\documentclass[pra,nofootinbib,twocolumn,amsmath,amssymb,superscriptaddress]{revtex4-2}
\usepackage{graphicx,amsmath,relsize,epstopdf,color,mathtools,bm,newtxtext,newtxmath,braket,rotating}
\usepackage[hyphenbreaks]{breakurl}
\usepackage[colorlinks=true,linkcolor=blue,citecolor=blue,urlcolor =blue]{hyperref}
\usepackage[normalem]{ulem}

\usepackage{booktabs}
\usepackage[table,xcdraw]{xcolor}

\usepackage{soul,xcolor}

\usepackage{dsfont}
\usepackage{xcolor}
\usepackage[inkscapelatex=false]{svg}
\usepackage[utf8]{inputenc}
\usepackage{lineno}

\begin{document}

\title{High-purity amplification of circularly polarized orbital angular momentum modes in an active spun ring-core tapered fiber}

\author{Iuliia Zalesskaia}
 \affiliation{Tampere University, Korkeakoulunkatu 3, Tampere, 33720, Finland}
 \email{iuliia.zalesskaia@tuni.fi}
 
\author{Hassan Asgharzadeh B.}
 \affiliation{Tampere University, Korkeakoulunkatu 3, Tampere, 33720, Finland}
 
\author{Zahra Eslami}
 \affiliation{Tampere University, Korkeakoulunkatu 3, Tampere, 33720, Finland}
 
\author{Hossein Fathi}
 \affiliation{Tampere University, Korkeakoulunkatu 3, Tampere, 33720, Finland}
 
\author{Evgenii Gribanov}
\affiliation{Ceramoptec SIA, Domes street 1a, Livani, LV-5316, Latvia}

\author{Andrey Grishchenko}
\affiliation{Ceramoptec SIA, Domes street 1a, Livani, LV-5316, Latvia}

\author{Florian Lindner}
\affiliation{Dept. Leibniz Institute of Photonic Technology e.V., Albert-Einstein-Str. 9, 07745 Jena, Germany
}%
\author{Katrin Wondraczek}
\affiliation{Dept. Leibniz Institute of Photonic Technology e.V., Albert-Einstein-Str. 9, 07745 Jena, Germany
}%

\author{Evgeny Savelyev}
\affiliation{Ceramoptec SIA, Domes street 1a, Livani, LV-5316, Latvia}

\author{Marco Ornigotti}
 \affiliation{Tampere University, Korkeakoulunkatu 3, Tampere, 33720, Finland}

\author{Valery Filippov}
\affiliation{Ampliconyx Ltd, Lautakatonkatu 18, Tampere, 33580, Finland
}

\author{Regina Gumenyuk}
 \affiliation{Tampere University, Korkeakoulunkatu 3, Tampere, 33720, Finland}
\affiliation{Ampliconyx Ltd, Lautakatonkatu 18, Tampere, 33580, Finland
}%

\begin{abstract}
Structured light, optical fields engineered in their spatial, polarization, or phase degrees of freedom, has become a key resource across advanced communication, sensing, imaging, and quantum technologies. Optical fibers nowadays play an essential role in this landscape, providing stable and scalable platforms for guiding, and amplifying complex modes such as vector and orbital angular momentum (OAM) beams. In this work, we demonstrate an active spun ring-shaped tapered fiber as a gain medium for efficient amplification of OAM modes preserving their modal purity and polarization topology. OAM beams with topological charges $\ell=1$ and $\ell=2$ carrying 60 ps pulses at 15 MHz repetition rate at 1030 nm wavelength are amplified over 1.2 W average power with modal purity over 95\%. The spatially resolved measurement of the OAM beam polarization topology revealed small distortion due to the coupling in to neighbour modes. These results demonstrate the high potential of active spun ring-shaped tapered fibers for power scaling of complex beams, preserving their phase and polarization structure simultaneously.
\end{abstract}

\maketitle

\section{Introduction}
Beams with a helical phase front of the type $\exp^{(i \ell \phi)}$, where $\ell$ is the topological charge of the beam, i.e., the number of 2$\pi$ phase shifts while winding around the beam's axis, are said to carry optical angular momentum (OAM) \cite{OAMlight}. Since their first discovery in optics \cite{berryNye, woerdman}, OAM beams have shown immense potential for application in several different areas of physics, ranging from particle manipulation \cite{partManipulation}, material processing \cite{Hnatovsky:10, PhysRevLett.110.143603}, nonlinear four-wave mixing \cite{nonlinearFWM1, nonlinearFWM2, nonlinearFWM3, nonlinearFWM4}, optical free-space and fiber-based communications \cite{OAMcomm}, super-resolution imaging \cite{superResImaging}, atomtronics \cite{atomtronics}, dense coding and data storage \cite{Bozinovic2013}, and quantum optics \cite{quantOptics}, to name a few. A comprehensive review and perspective on the field can be found, for example, in Ref. \cite{OAMroadmap}.
%
%around the center of the phase profile of the beam carrying optical orbital angular momentum (OAM), have immense potential for applications in 
%
%optical trapping, laser material processing, plasma fusion, quantum information and optical communications\cite{OAMlight, OAMcomm, OAMforbes}. Benefit from its unique properties, the OAM as a new degree of freedom of light has attracted much attention in various application scenarios, such as optical trapping and manipulation of particle \cite{partManipulation}, material processing [.], nonlinear four-wave mixing [.], optical free space and fiber-based communications \cite{OAMcomm}, super-resolution imaging [10], atomtronics [.], ultrahigh-density data storage \cite{OAMdense}, and other fields. 
%
Current research in OAM spans the generation, detection, manipulation, amplification, and transmission of OAM beams, with the majority of work to date limited primarily to bulk optics applications. OAM beams, in fact, are typically generated by imposing a helical phase front on a free-space-propagating Gaussian beam using, for example, a spatial light modulator (SLM) \cite{SLMgeneration}, a phase plate \cite{OAMlight}, or q-plate \cite{qPlate} quite often imposing parasitic radial modes. On the other hand,  the versatility and potential in power and dimensional scalability of OAM beams in optical fibers have recently attracted a lot of attention from the scientific community.
%efficient power of OAM mode in the fiber is limited by 
%
%low conversion and coupling efficiencies, and by device damage thresholds. To address these issues, optical amplification in fiber could be a potentially effective solution.  

Over the past decade, rare-earth-doped fiber lasers and amplifiers have seen rapid development owing to excellent beam quality, efficiency, and scalability. Especially, significant progress has been achieved in Yb-doped large mode area (LMA) fibers driven by the improved homogeneity across the large core and cladding and more precise control of the fiber fabrication process  \cite{Jain:15, Limpert2007, fathi2025versatile}. The enlarged mode area of optical fibers enabled dramatic power scaling of fundamental mode, far breaking the kW power level  with excellent beam quality \cite{Sun:22}. Yb-doped LMA fibers were found to be suitable for the amplification of not only Gaussian beams but also radially or azimuthally polarized beams exceeding the hectawatt level or milli-Joule energy of nanosecond pulses \cite{Lin:17, Liu:24, Koyama:11pico}. Despite significant advances in power level, the parasitic amplification of the fundamental mode degraded the quality of beams with complex polarization  or phase structures.

Later, theoretical analysis proved that OAM modes can be eigenmodes of specially designed fibers \cite{RamachandranKristensen2013}. Although first attempts for amplification of low-order OAM modes have been demonstrated even in Yb-doped LMA fibers, reaching several tens of watt power level \cite{Koyama:11pico, Koyama:11nano}, to maintain the stable propagation and robustness against external perturbation it is essential that the index separation between adjacent vector modes are greatly enhanced (to $>\Delta  n_{eff}10^{-4}$), suppressing inter-modal coupling \cite{Bozinovic2013}. To achieve this condition, several types of fiber architectures were engineered, such as ring-core and twisted fibers \cite{Wu:22, Russell2017, Bozinovic2013, Gregg:15, Liu2022Nature}. The fiber twisting (with few millimeters or less) induces circular internal birefringence, which results in different propagation constants quenching mode degeneracy and maintaining not only generation but also stable propagation of OAM modes in the fiber \cite{Russell2017}. Subsequent works confirmed that ring- or air-core fibers are also capable of lifting mode degeneracy and preserving OAM structure under external perturbations, enabling long-distance transmission and amplification of high-order OAM \cite{Wen:23, Brunet:14}. The CW signal amplification in Yb-doped ring-core fiber demonstrated more than 8 dB gain for 20 OAM modes with differential mode gain less than 1 dB \cite{Ou:22}. Although the structure of OAM modes was well preserved in the ring-core fiber, the power level was limited by the moderate gain.

Recently, an active spun tapered double-clad fiber (sT-DCF) was proposed as an alternative approach for power scaling of short pulsed signals with complex polarization structures \cite{Zalesskaia:24}. Featuring circular polarization eigenstates, the Yb-doped sT-DCF preserved the spatial and polarization profiles of a cylindrical vector beam with 10 ps pulses along the whole power range up to 22 W, while a parabolic longitudinal taper raises nonlinear thresholds without sacrificing modal integrity. This fiber design incorporated the circular core, supporting the undesired fundamental mode amplification in the central part of the core and resulting in the growing beam deterioration, which limited the power level.

In this work, we demonstrate an Yb-doped ring-shaped core spun tapered double-clad fiber as a gain medium for amplification of OAM modes while preserving their complex profile in phase and polarization domains. The Yb-doped ring-shaped core sT-DCF (sT-RCF) is designed to support and amplify two OAM modes at 1030 nm, based on the specially developed theoretical framework that incorporates both twisting and longitudinal fiber architectures, together with its active nature. OAM-carrying 60 ps pulses at 15 MHz repetition rate are free-space coupled into sT-RCF and amplified over 1.2 W average power with the gain exceeding 11.5 dB. The comprehensive experimental analysis of phase and polarization topology of the amplified OAM modes supported by the simulation results confirms the excellent preservation of beam profiles in both domains simultaneously, revealing the modal purity of 95\% and 97\% for OAM1 and OAM2, respectively. These results open the route to high-power amplification of OAM modes with precise control of the complex beam patterns in multiple domains.

\section{Theory}
The longitudinal variation in the refractive index profile induced on an optical fiber by twisting requires, in general, a full 3D approach for its modal analysis, since the position of the transverse refractive index profile of the fiber varies along the propagation direction, following the twisting profile. To avoid such computationally intense problem, however, we employ transformation optics \cite{trafoOptics, PCFbook} to map a twisted optical fiber in the laboratory frame into a straight fiber, as seen from a reference frame solidal with the twisting itself. In doing so, the fiber appears straight, and effect of twisting is mapped into anisotropic permittivity and permeability \cite{HRMarxiv,PCFbook}. In such a transformed reference frame, moreover, the OAM modes emerge as the natural set of eigenmodes for the fiber, and can be manipulated directly. Consequently, the general form of the coupled mode equations (CMEs) for the various modes of the fiber in the twisting frame can be casted in the following form \cite{HRMarxiv}
%
%by twisting a fiber makes it necessary to employ a full 3D approach for the modal analysis of the structure. However, the application of transformation optics formalism significantly reduced this computational complexity by mapping the evolution of light in a twisted fiber to an equivalent straight fiber, where the effect of twisting is mapped into anisotropic permittivity and permeability \cite{HRMarxiv, trafoOptics, PCFbook}. In such a transformed reference frame, the OAM modes emerge as the natural eigenmodes of the twisted fiber \cite{HRMarxiv}. Based on this, a complete coupled mode theory can be developed, to account for
%As demonstrated in Ref. \cite{HRMarxiv}, vector Orbital Angular Momentum (OAM) modes are, in fact, true natural eigenmodes of twisted optical fibers. In addition, a rigorous theoretical framework based on Coupled Mode Theory (CMT) has been established to address the propgation characteristics in such fibers [1]. Consequently, the general form of Coupled Mode Equations (CMEs) for twisted fibers can be expressed as
\begin{equation}
\frac{da_j(z)}{dz}
= \sum_{q} a_q(z)
\left\{
i K_{jq}\,
e^{i\left[\Delta\beta_{qj} \pm \left(J_q - J_j\right)\Upsilon\right]z}
+ \Omega_{jq}\, e^{i\Delta\beta_{qj}z}
\right\},
\label{eq:coupled_mode}
\end{equation}
where $a_j$ is the slowly-varying mode amplitude, $\Upsilon$ is the twisting rate, $\Delta\beta_{qj}=\beta_q-\beta_j$ is the propagation constant mismatch between the modes $q$ and $j$, while $J_q$ is the total angular momentum carried by mode $q$. Moreover, $K_{jq}$ is the twisting-induced coupling coefficient, while $\Omega_{jq}$ is the taper-induced coupling coefficient between modes $j$ and $q$. The coupling coefficient $K_{jq}$ is then calculated with reference to an untwisted optical fiber in the transformed reference frame, and its explicit expression is given by \cite{HRMarxiv}
\begin{equation}
K_{jq}
= \frac{k_0}{4}\sqrt{\frac{\varepsilon_0}{\mu_0}}
\iint
\mathbf{E}_q^*\left(r,\varphi\right)
\cdot
\delta\boldsymbol\varepsilon_{qj}
\cdot
\mathbf{E}_j\left(r,\varphi\right)\,
r \, dr \, d\varphi,
\end{equation}
where $k_0$ is the free-space wavenumber, $\mathbf{E}_q\left(r,\varphi\right)$ and $\mathbf{E}_j\left(r,\varphi\right)$ are the normalized transverse electric field modes of the reference, untwisted fiber. The tensor $\delta\boldsymbol\varepsilon_{qj}$ then represents the perturbation induced by the twisting to the permittivity of the fiber. Notice, moreover, that the integration above is carried out in the twisted frame, solidal to the fiber twisting. 
%of the reference modes $j$ and $q$, respectively. The term $\left[\delta\varepsilon\right]$ represents the induced perturbation in the permittivity and, in general, can be expressed as a tensor in twisted fibers. The constants $k_0$, $\epsilon_0$, and $\mu_0$ are the free-space wavenumber, permittivity, and permeability, respectively. The superscript $(\prime)$ indicates that the radial coordinate $r'$ and the azimuthal coordinate $\varphi'$ are expressed in the helical coordinate frame [1]. 
%
To extend this analysis to fiber amplifiers, a term like $g(z)a_q(z)/2$, accounting for the gain of the active medium, must be incorporated into the right-hand-side of Eq.~\ref{eq:coupled_mode} to accurately model the propagation dynamics of the signal and pump light %\cite{HRMinProgress}. 

\begin{figure}[ht]
\centering
\includegraphics[width=\linewidth]{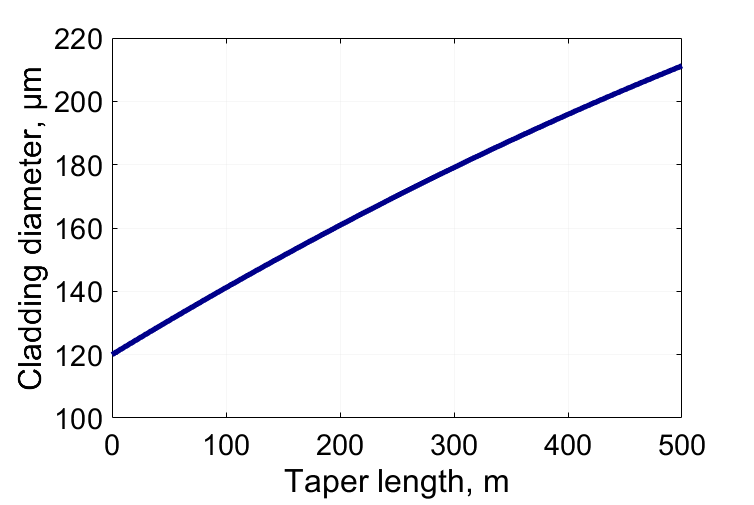}
\caption{Theoretically calculated longitudinal cladding-diameter profile of the sT-RCF. }
\label{fig:theoryprofile}
\end{figure}

Figure \ref{fig:theoryprofile} illustrates the theoretically calculated longitudinal profile of the sT-RCF cladding diameter, with its functional form and parameters detailed in Table \ref{tab:parameteres}. This profile ensures a gradual variation of the fiber radius along the propagation coordinate z.  As a result, the magnitude of the taper-induced coupling coefficients, $\Omega_{jq}$, is minimized. 

%Fig. \ref{fig:resultModel} illustrates the intensity and phase distribution of the output beam of a Yb-doped, tapered, twisted ring-core fiber amplifier. All simulation parameters are listed in Table \ref{tab:parameteres} .  Furthermore, a macro-bend, with a bending radius of 20 cm, is assumed in simulation. With this, bending acts as a source of perturbation, introducing an asymmetry that modifies the permittivity of the reference fiber. 

\begin{table}[htbp]
\centering
\caption{Simulation parameters used to generate the numerical results presented in Fig. \ref{fig:resultModel}.}
\scriptsize
\resizebox{\linewidth}{!}{%
\begin{tabular}{|c|c|c|c|}
\hline
$n_{co}$ & 1.4543 &
Wider to smaller ratio (tapered fiber), $\Pi$ & 1.76 \\
\hline
$n_{cl}$ & 1.4503 &
Fiber length $L$ [m] (Case: $\mathrm{OAM}_{1}$) &
$\approx 4.6$ \\
\hline
$\lambda_{\text{sig}}$ [nm] & 1039 &
Fiber length $L$ [m] (Case: $\mathrm{OAM}_{2}$) &
$\approx 2.9$ \\
\hline
$\lambda_{p}$ [nm] & 976 &
Tapering profile &
$\displaystyle R(z)=\frac{b_{0}-b_{f}}{2L}z^{2}+\frac{b_{f}}{2}+2R^{s}$ \\
\hline
\begin{tabular}[c]{@{}l@{}}Inner radius of the core\\
(narrower side), $R_{1}^{s}$ [$\mu$m]\\
(Case: $\mathrm{OAM}_{1}$)\end{tabular}
& 1.50 &
Average taper angle, $b_{0}$ &
$\displaystyle 2\frac{R^{w}-R^{s}}{L}$ \\
\hline
\begin{tabular}[c]{@{}l@{}}Outer radius of the core\\
(narrower side), $R_{2}^{s}$ [$\mu$m]\\
(Case: $\mathrm{OAM}_{1}$)\end{tabular}
& $3\times R_{1}^{s}$ &
Taper shape factor, $b_{f}$ &
1.2 \\
\hline
\begin{tabular}[c]{@{}l@{}}Inner radius of the core\\
(narrower side), $R_{1}^{s}$ [$\mu$m]\\
(Case: $\mathrm{OAM}_{2}$)\end{tabular}
& 1.910 &
Twist pitch [mm] &
3.6 \\
\hline
\begin{tabular}[c]{@{}l@{}}Outer radius of the core\\
(narrower side), $R_{2}^{s}$ [$\mu$m]\\
(Case: $\mathrm{OAM}_{2}$)\end{tabular}
& $3\times R_{1}^{s}$ &
Pump power [W] &
\begin{tabular}[c]{@{}l@{}}
Higher power $\mathrm{OAM}_{1}$: 3.6\\
Lower power $\mathrm{OAM}_{1}$: 1.3\\
Higher power $\mathrm{OAM}_{2}$: 4.0\\
Lower power $\mathrm{OAM}_{2}$: 1.3
\end{tabular} \\
\hline
\end{tabular}%
\label{tab:parameteres}
}
\end{table}

\section{Materials and methods}
\subsection{Double-clad tapered fiber with a ring-shape active core}

 %sT-RCF was manufactured by using a ring shaped active core preform with CCDR 3. The Yb-doped ring produced by REPUSIL technology was filled with pure silica. A cross-sectional refractive index and Yb2O3 concentration profile of the initial core material is illustrated in Figure \ref{fig:profile} (Top left inset). At a wavelength of 976 nm, the in-core absorption coefficient of the ring shape core material preform was measured to be  ?660? dB/m, while the numerical aperture was determined to be 0.125.
 %The final taper preform was constructed with a core-cladding-diameter ratio (CCDR) equal to 14.4. In order to achieve the desired fiber tapering, a continuous variation of the fiber drawing speed was applied. Additionally, for the sT-RCF, a constant preform rotation velocity was implemented during the drawing process to impart the spun architecture. The fibers underwent a tapering process wherein the outer cladding diameter was gradually reduced along a 5.15 m length, resulting in a final cladding diameter range of 115 µm to 213 µm.

\begin{figure}[ht]
\centering
\includegraphics[width=\linewidth]{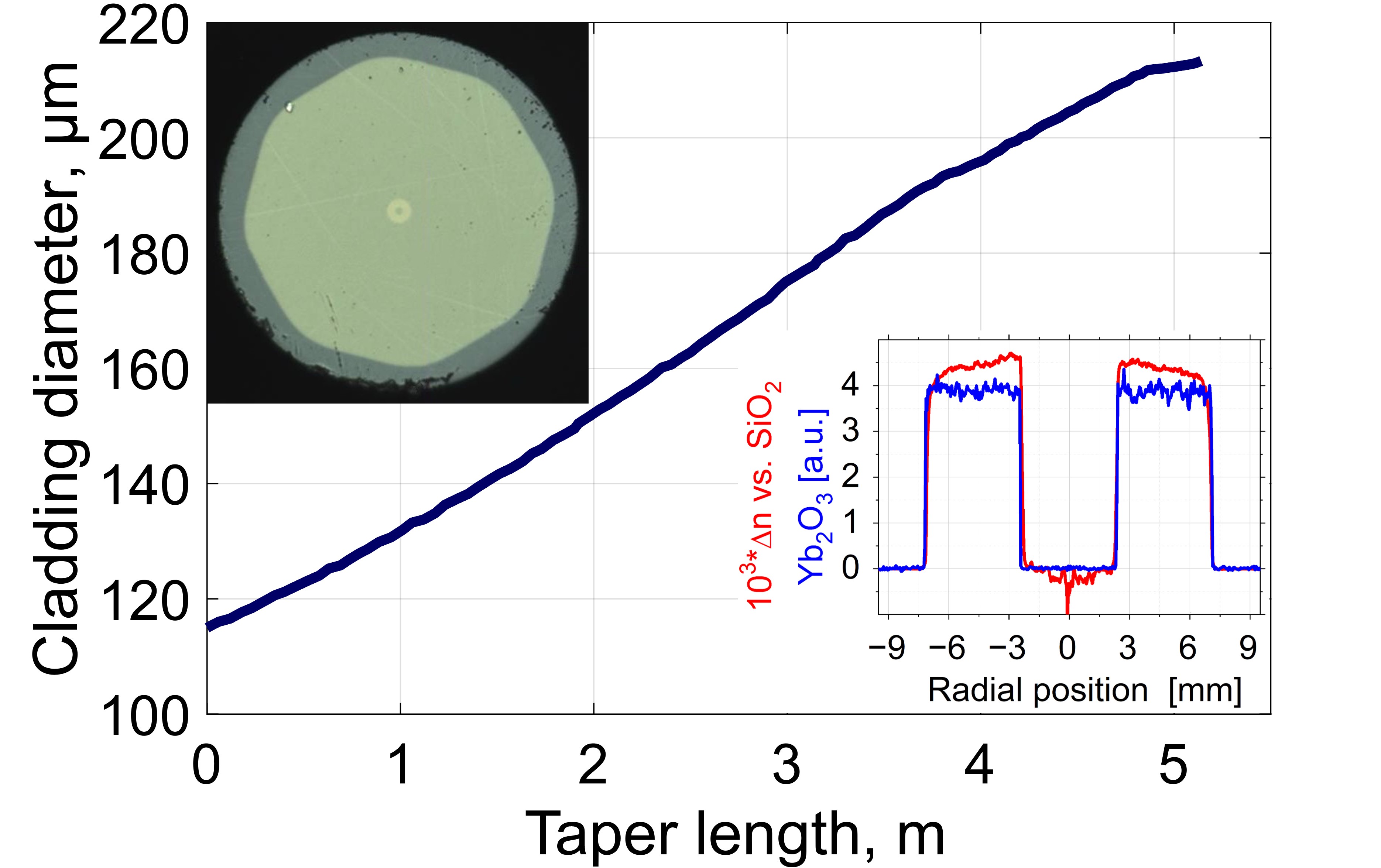}
\caption{The longitudinal cladding-diameter profile of the sT-RCF. Top left inset: fiber facet at the wide side. Bottom right inset: refractive-index and Yb\textsubscript{2}O\textsubscript{3} concentration profile of the initial core material.}
\label{fig:profile}
\end{figure}

 Manufacturing of a double-clad tapered fiber with a ring-shape active core involved three main steps: (1) synthesis of Yb\textsubscript{2}O\textsubscript{3}-doped silica glass rod with pure silica center  (initial core preform), (2) adaptation of preform geometry (final preform), and (3) taper drawing.
 The core preform was fabricated by REPUSIL method and was a monolithic silica glass rod with an Yb\textsubscript{2}O\textsubscript{3}-doped ring-shaped core. We employed proprietary protocols for doping a high purity silica suspension (Heraeus Quarzglas GmbH \& Co. KG) using water soluble salts of dopant precursors, and for transforming it into a vitrified core preform rod containing a pure silica center (F300, Heraeus Quarzglas GmbH \& Co. KG).  AlCl\textsubscript{3}*6H\textsubscript{2}O (99.999\%, Sigma Aldrich) and YbCl\textsubscript{3}*6H\textsubscript{2}O (99.99\%; Auer Remy) served as water soluble oxide precursors. pH-adjusted doping solutions were prepared in ultrapure water (GPR Rectapur, VWR Avantor) adjusted with HCl (32\%, EMSURE for Analysis, Merck) or NH\textsubscript{3} (25\%, Suprapur, Merck).  An example of a cross-sectional refractive index and Yb\textsubscript{2}O\textsubscript{3} concentration profile of the initial as-vitrified core material is illustrated in Fig. \ref{fig:profile} (bottom right inset). The as-obtained doped part had a numerical aperture of 0.113.
 
The ratio of the diameter of the undoped part of the core to its total diameter (outer diameter of doped part) was 1/3. The absorption of the doped region of the core material at a wavelength of 976 nm was 880 dB/m.
 One of the main advantages of the REPUSIL method is the ability to produce preforms with a large absolute core size, which subsequently allows the production of a large amount of active fiber based on it. However, before using such material for active tapers drawing, it is necessary to ensure the required ratio between the first cladding and the core of the light guide. To increase the initial cladding to core diameter ratio (CCDR) from $\approx$1.4, the core preform was sleeved using a silica tube, followed by elongation of the formed monolithic rod. In the next step, to increase the pump absorption and make it more uniform along the entire length of the active double-clad optical fiber, the symmetry of the first cladding was reduced by changing the shape of the first cladding from a circle to a regular octagon (Fig.\ref{fig:profile}, top left inset). The flat-to-flat octagonal cladding size to core diameter ratio after these operations became 12. Since our sT-RCF is designed for use in high-power amplifiers, we deposited second silica cladding, doped with fluorine. The first cladding numerical aperture was 0.27 and the preform outer diameter to core diameter ratio was 14.4.
The preform prepared in the above-described way was placed in a special drawing tower, which, during the fiber drawing procedure, allows to change independently three parameters: the preform feed rate into the high-temperature furnace, its rotation frequency, and the drawing speed. All these parameters were changed according to a predetermined algorithm, which made it possible to manufacture the desired longitudinal profile of sT-RCF with a constant pitch. The total length of the sT-RCF was 5.14 m, with an outer cladding diameter of 115 µm and 213 µm for the narrow and wide ends, respectively (shown in Fig. \ref{fig:profile}), measured pitch was 3.67 mm.

\subsection{Experimental setup}
\begin{figure*}[ht]
\centering
\includegraphics[width=\textwidth]{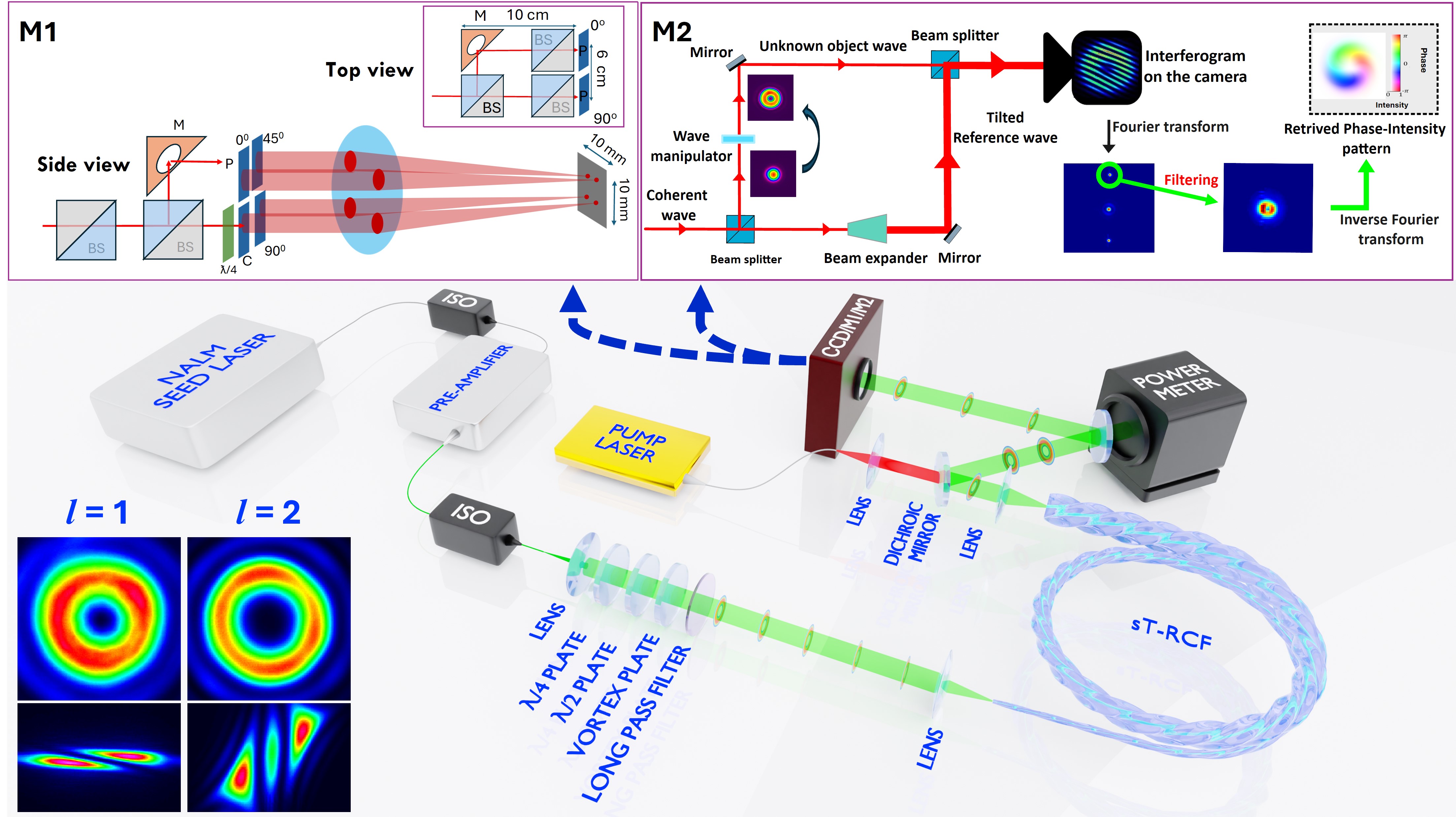}
\caption{Schematic of the sT-RCF amplifier experimental setup and beam diagnostic modules. Top left (M1): polarimetry setup. Top right (M2): digital off-axis holography setup and phase reconstruction. Bottom left: $\ell=1$ (OAM1) and $\ell=2$ (OAM2) beams generated by vortex plates and their intensity modes at the focal point of a cylindrical lens.}
\label{fig:setup}
\end{figure*}

In the experimental configuration shown in the Fig. \ref{fig:setup}, a mode-locked seed laser based on a nonlinear amplifying loop mirror (NALM) generated linearly polarized pulses at a 15 MHz repetition frequency, which were consequently stretched up to 60 ps to avoid the raise of nonlinearities in the 
pre-amplifier stage as the limiting factor to achieve the desired power level. 

The pulses were guided into a Yb-doped polarization-maintaining (PM) single-mode fiber pre-amplifier, yielding an average output power of 330 mW. The amplified beam was then sent through $\lambda$/4 and $\lambda$/2 plates to obtain a left-handed circularly polarized beam, and subsequently directed to a vortex phase plate (VPP) (V-1030-10-1 and V-1030-10-2, Vortex Lens for 1030 nm and topological charges 1 and 2, respectively, Vortex Photonics, Germany), where a vortex beam was formed. With a diameter of 10 mm, the VPP poses 64 steps, operating at 1030 nm. 
The transformed vortex beam exhibited a prominent doughnut-shaped intensity profile in the far field, as depicted in the bottom-left inset in Fig.\ref{fig:setup}. Also, the beam after the VPP was analyzed using a cylindrical-lens technique, which showed the clear transformation of a Gaussian beam into a vortex beam with topological charges equal to 1 and 2, respectively. The vortex beam was then coupled into the high-power amplifier based on the sT-RCF, yielding coupling efficiencies of 26$\%$ for $\ell=1$ (OAM1) and 24 $\%$ for $\ell=2$ (OAM2). The sT-RCF was counter-pumped with a fiber-coupled 976 nm high-power wavelength-stabilized diode laser (BWT K976AG1RN), pigtailed with a 105 µm core fiber (NA = 0.22). To protect the pre-amplifier from back-propagating unabsorbed pump light coming from the tapered fiber, a long-pass filter was inserted into the vortex beam path.
 
 To ensure the undistorted propagation of the OAM1 and OAM2 modes in the fiber, the sT-RCF was cut to core/cladding diameters from the narrow side to 9/130 µm for the amplification of OAM1 and to 11.8/170 µm for the amplification of OAM2 mode. The wide side of the tapered fiber remained the same. The sT-RCF length was 4.6 m and 2.9 m for OAM1 and OAM2 modes amplification, respectively. The cladding diameter fiber profile is shown in Fig. \ref{fig:profile}. The spectrum of the amplified signal was measured using an optical spectrum analyzer (Yokogawa AQ6374E) and presented in Fig.\ref{fig:resultExp}. The output beam after amplification in the sT-RCF was characterized using several complementary diagnostics to verify the intensity, phase and polarization distributions, and mode content.

\subsection{Modal analysis}
The far-field intensity profile was recorded with a beam profiler (Spiricon Ophir BGP-USB3-LT665). For an initial assessment of the modal structure, a cylindrical-lens technique was employed by inserting a Plano-Convex cylindrical lens (Thorlabs LJ1629RM) into the beam path before and after the sT-RCF amplifier \cite{VAITY20131154}.
To perform a full-field analysis of the amplified beam, we employed off-axis digital holography \cite{Goodman1967, Off-axis:11}. A fraction of the amplified output beam reflected by a beam sampler was used as the unknown object wave. The reference wave was derived from the NALM seed laser by tapping a small fraction of the beam using a 90/10 coupler and sending it through a compensation line (not marked in the schematic of the experimental setup) to equalize the optical path lengths. The off-axis interferometer was formed by introducing a small angle between the object and reference waves, and the resulting hologram was recorded on the beam profiler.

%(Ophir SP90378, 4.54 \(\mu \)m effective pixel pitch)

 \subsection{Polarization Topology}
The polarization state of the optical beam was characterized using a division-of-amplitude Stokes polarimeter, enabling single-shot reconstruction of the full Stokes vector with spatial resolution across the beam profile \cite{Moller:23}. A schematic of the compact, two-folded Stokes polarimeter is shown in the inset (M1) of the Fig.\ref{fig:setup}. The collimated input laser beam is divided into four parallel detection channels using non-polarizing beam splitters. Each channel projects the incident field onto a distinct polarization state by means of linear polarizers oriented at 0°, 90°, and 45°, and a quarter-wave plate followed by a linear polarizer for circular polarization analysis.

The resulting intensity distributions, denoted \( I_0 \), \( I_{45} \), \( I_{90} \), and \( I_C \), corresponding respectively to linear polarization at 0°, 45 and 90°, and circular polarization, were recorded simultaneously on a camera. These measurements allow reconstruction of the polarization state of the incident light by computing the four-component Stokes vector through a linear matrix relation,
\begin{equation}
\mathbf{S} = \mathbf{M}\,\mathbf{I},
\end{equation}
where
\begin{equation}
\mathbf{S} =
\begin{pmatrix}
S_0 \\ S_1 \\ S_2 \\ S_3
\end{pmatrix},
\qquad
\mathbf{I} =
\begin{pmatrix}
I_0 \\ I_{90} \\ I_{45} \\ I_C
\end{pmatrix}.
\end{equation}

In an ideal system, the reconstruction matrix \( \mathbf{M} \) takes the form
\begin{equation}
\mathbf{M} =
\begin{pmatrix}
1 & 1 & 0 & 0 \\
1 & -1 & 0 & 0 \\
-1 & -1 & 2 & 0 \\
-1 & -1 & 0 & 2
\end{pmatrix}.
\end{equation}
In practice, however, experimental non-idealities, including unequal transmission between channels, imperfect optical components, and the camera response—cause deviations from this ideal matrix, making calibration necessary.

Calibration was performed by generating a set of at least four known input polarization states using a combination of half-wave and quarter-wave plates placed before the polarimeter. For each calibration state, the corresponding intensity signals from the four detection channels were recorded, forming a measurement dataset relating the known input Stokes vectors to the detected intensities. From this dataset, a calibrated instrument matrix was determined, incorporating all system non-idealities. This calibrated matrix was subsequently used to reconstruct the Stokes vector of unknown input beams.

%\begin{figure*}[t]
%\centering
%\includegraphics[width=\textwidth]{Polarization_elipses.png}
%\caption{Spatially resolved polarization ellipses for OAM1 (a-b) and OAM2 (c-d) beams at low and high amplified powers %respectively.}
%\label{fig:polarization}
%\end{figure*}

For each measurement, the recorded intensity images were first background-corrected and spatially registered. The Stokes parameters were then calculated on a pixel-by-pixel basis, with a spatial resolution of 20 µm, by applying the calibrated instrument matrix to the measured intensity vector at each spatial position.

\section{Results}

\subsection{Phase, modal and spectrum characterization of the amplified OAM beams}

\begin{figure*}[!t]
\centering
\includegraphics[width=\textwidth]{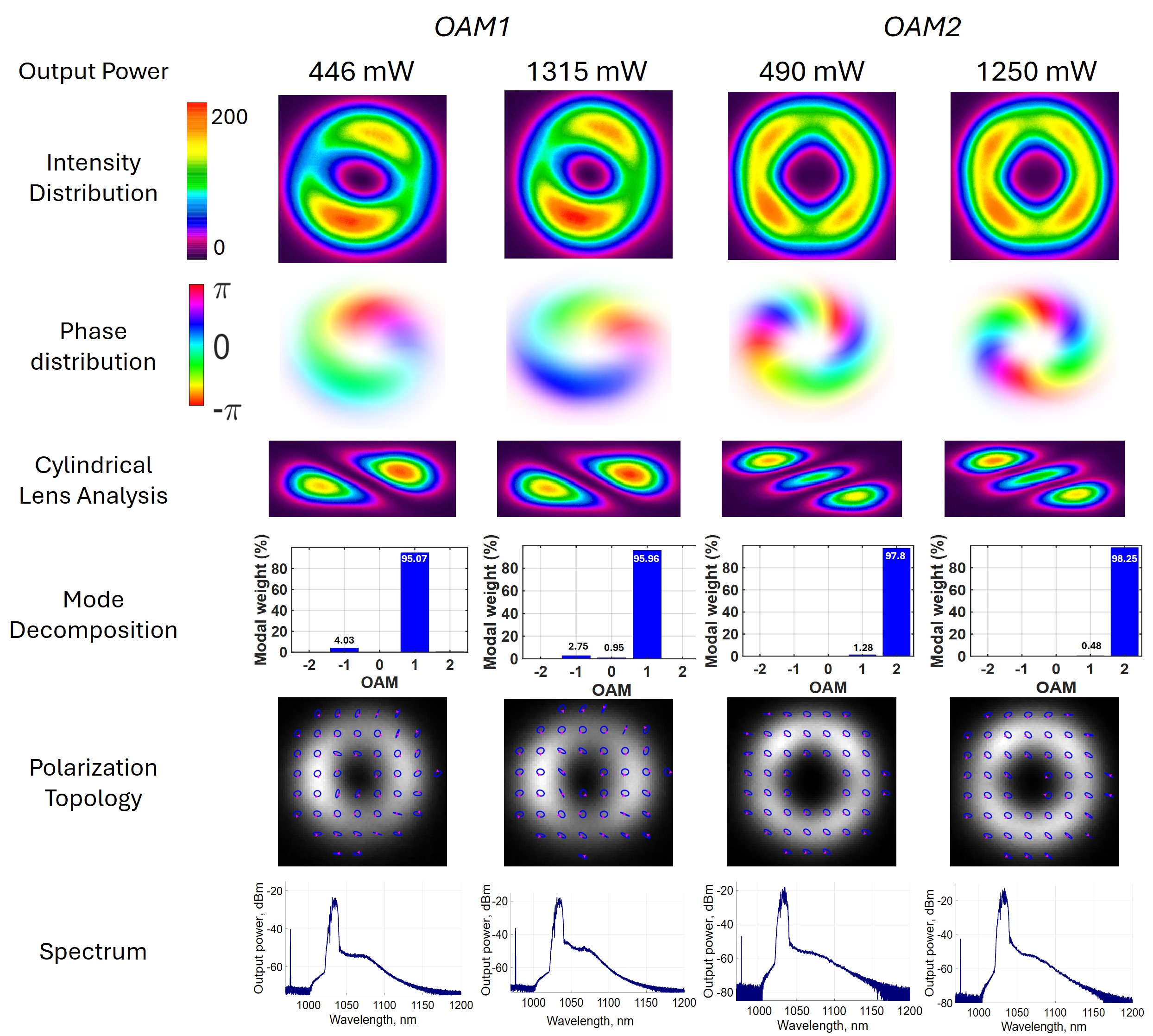}
\caption{Experimental phase, modal and spectral characterization of the amplified OAM beams at the sT-RCF output. Measured intensity and phase distributions, cylindrical-lens (astigmatic) patterns, mode decomposition, polarization maps and optical spectrum for OAM1 and OAM2 at two output power levels (446 mW and 1315 mW for OAM1; 490 mW and 1250 mW for OAM2).}
\label{fig:resultExp}
\end{figure*}

In the first experiment, the OAM1 mode was coupled into the sT-RCF. The fiber length was approximately 4.6 m, with a core diameter of 9 µm at the thin input side and 14.9 µm at the opposing side. This input sT-RCF geometry supported propagation of the OAM1 mode, while the output side of the tapered fiber supported both OAM1 and OAM2 modes. To minimize perturbation along the mode propagation, the fiber was loosely coiled with a large bend diameter of around 30 cm. The thin end of the sT-RCF was cleaved perpendicular to the fiber axis to optimize the vortex beam injection efficiency and avoid beam distortion. The experimentally measured coupling efficiency in this configuration was 26 $\%$, corresponding to 85 mW input power at 1030 nm.  By contrast, the output facet was angle-cleaved to suppress back reflections. The final amplifier stage achieved an optical-to-optical efficiency of 36$\%$ for OAM1 mode amplification (from the total pump power to the sT-RCF output power), delivering 1.3 W, which corresponded to an amplification level of 11.87 dB.

The experimentally measured spatial intensity distribution, phase distribution, cylindrical lens analysis, mode decomposition, and spectrum of amplified OAM1 mode are shown in Fig. \ref{fig:resultExp}. The measurements were performed at two power levels: the minimum output power, limited by the pump diode operation, and the maximum output power, limited by the onset of parasitic lasing on the spontaneous emission spectrum. For the output power measurement, the residual pump peak was filtered and remained below 0.5$\%$ of the total output power. The minimum output power of 446 mW was obtained with 1.3 W of pump power at 976 nm. The amplified output beam exhibited a doughnut-shaped intensity distribution with two prominent side lobes. The measured spectrum showed a signal-to-ASE contrast exceeding 25 dB.
The cylindrical lens analysis revealed a clear two-lobe pattern, confirming the strong content of OAM1 mode in the amplified beam. The phase distribution was reconstructed from the recorded interference pattern using off-axis digital holography setup, showing a 2$\pi$ phase change per pass. The mode purity of the amplified beams was evaluated by computing the overlap between OAM modes of topological charge $\ell$ and the corresponding spiral phase factor $\exp(i \ell \phi)$. Analysis of the same off-axis hologram indicated an OAM1 mode content of over 95\% in the amplified beam. The remaining 5\% of the power was coupled to $\ell=-1$, while during amplification this power fraction was further distributed between $\ell=-1$ and the fundamental mode, keeping the power fraction of the desired OAM1 mode nearly unchanged.
The spatially resolved polarization maps across the beam were derived from the reconstructed Stokes parameters, as shown in Fig.\ref{fig:resultExp}. The measurements demonstrate that the initially circularly polarized beam remains largely preserved after propagation and amplification in the tapered fiber, with only a slight increase in ellipticity.

%The mode purity of the amplified beams was evaluated by computing the overlap between optical vortices of topological charge $\ell$ and the corresponding spiral phase factor $\exp(i \ell \phi)$.

To experimentally examine the preservation and amplification of OAM2 mode, the sT-RCF was shortened to 2.9 m having a core diameter of 11.8 µm at the thin input side. The output diameter of the tapered fiber remained the same. The experimentally measured coupling efficiency decreased to 24$\%$, resulting in 79 mW input power at 1030 nm. The amplification of OAM2 mode reached 12 dB delivering 1.25 W output power at 4 W pump power, while the optical-to-optical efficiency reached 31.25$\%$.

The same beam characterization methods were applied to qualify the OAM2 amplification, and the corresponding results are shown for comparison in Fig. \ref{fig:resultExp}. Measurements were also performed at the lowest stable pump power allowed by the pump diode. The minimum output power of 490 mW was obtained at 1.3 W of pump power.
The amplified output beam maintained a doughnut-shaped intensity distribution with four prominent maxima, and the measured spectrum exhibited a signal-to-ASE contrast exceeding 25 dB, as in the previous experiment.
The cylindrical lens analysis produced a clear three-lobe pattern, confirming the strong content of the OAM2 mode in the amplified beam. The phase distribution was retrieved from the recorded interference pattern using digital off-axis holography, showing a 4$\pi$ phase change per pass, and mode decomposition derived from the same data yielded an OAM2 mode content exceeding 97\% in the amplified beam. The residual power fraction of 3\% was coupled into OAM1 mode, which during amplification decreased due to small coupling into OAM2 mode and OAM3 mode. Though, theoretically, the fiber design does not support higher than OAM2 modes, the bending-induced coupling at the thick part of the tapered fiber could potentially create the condition for coupling of small power fractions into higher order modes. 

\begin{figure*}[!t]
\centering
\includegraphics[width=\linewidth]{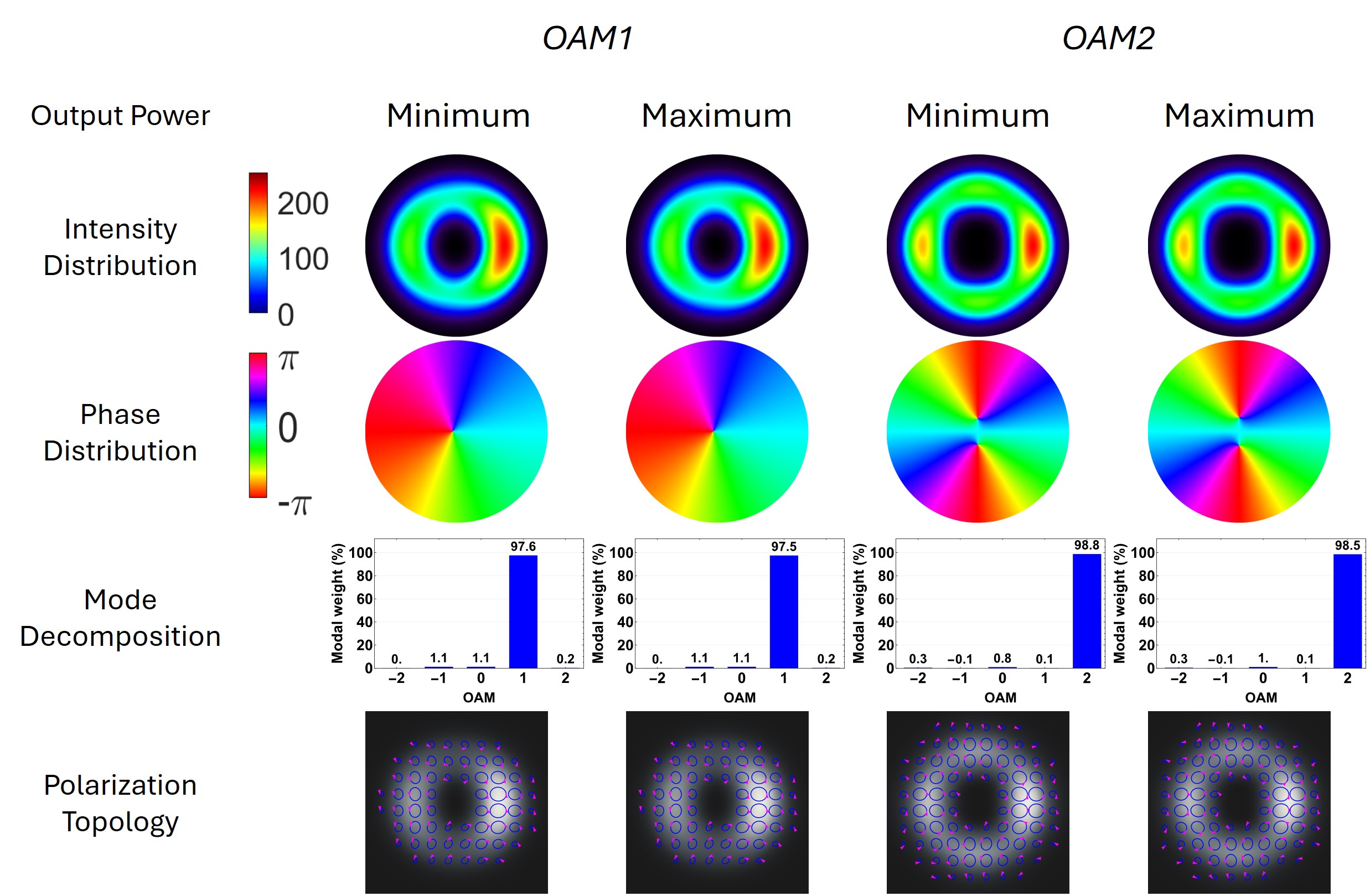}
\caption{Simulation data obtained for the amplified OAM beam at the sT-RCF output. The output power, intensity and phase distributions, mode decomposition, and polarization maps corresponding to OAM1 and OAM2 are presented in rows 2–5, respectively. For each OAM value, low (left column) and high (right column) output power data are presented, mirroring the structure of Fig. \ref{fig:resultExp}. The values for the simulation parameters are the same as those listed in Table \ref{tab:parameteres}.}
\label{fig:resultModel}
\end{figure*}

Spatially resolved polarization maps further indicate that the initially circularly polarized beam remains largely preserved after propagation through the spun fiber, with only a slight increase in ellipticity.

In both experiments, efficient coupling into the sT-RCF and stable amplification of the targeted OAM modes were successfully realized. The doughnut-shaped intensity profile, the cylindrical-lens patterns, and the phase retrieved by digital off-axis holography confirm robust maintenance of high modal purity for both modes. Specifically, digital off-axis holography shows the expected 
2$\pi$ and 4$\pi$ phase changes for OAM1 and OAM2 modes, respectively. The astigmatic transform produced the corresponding number of lobes in the Fourier plane of the cylindrical lens for the coupled OAM modes providing another confirmation of phase pattern conservation during the amplification process.
\subsection{Theoretical modeling results}

Figure \ref{fig:resultModel} depicts the output power, intensity and phase distribution, mode decomposition, and polarization maps for the amplified OAM1 and OAM2 modes. For each mode, the left and right columns correspond to the low and high output power regimes, respectively. The input seed powers were fixed at 85 mW for OAM1 and 79 mW for OAM2 matching experimental results. While mode decomposition indicated that OAM1 and OAM2 capture more than 97.5\% of the output power (third row), the intensity distributions deviated from a perfect doughnut shape, typical of OAM-carrying beams \cite{OAMlight}, exhibiting instead a number of lobes compatible with the OAM carried by each mode, i.e., $2\ell$ lobes for a mode carrying $\ell$ units of OAM. According to the simulations, this was due to the fact that the output mode was not purely an OAM mode, but rather a superposition of different modes, most importantly of modes with opposite OAM, than the desired one (see, for example, the mode decomposition panel in Fig. \ref{fig:resultModel}). Mutual coupling between these modes happened because of bending, as well known in the literature \cite{Ulrich:80}. This perturbation broke the degeneracy between orthogonal linear polarization states, making so that $x$- and $y$-polarized modes propagated at different speeds in the fiber. When transformed into the circular polarization basis, this manifested as a weak coupling term between left- and right-circular polarization states, enabling coupling between OAM modes of opposite charge which, in turn, gave rise to the lobed pattern of the output mode \cite{OAMlight}. This bending-induced coupling was visible in our numerical result in the polarization map (see the corresponding row in Fig. \ref{fig:resultModel}), where the presence of local elliptical polarization across the mode profile corroborated the nature of this interaction. Similar local imperfections were visible on the experimental polarization measurements.
%
%Moreover, it is well known that bending induces linear birefringence  \cite{Ulrich:80}, breaking the degeneracy between orthogonal linear polarization states. This implies that modes with $x$- and $y$-polarizations propagate at distinct speeds. When transformed into the circular polarization basis, this manifests as a weak coupling term between left- and right-circular polarization states. we incorporate this coupling into our theoretical model. The resulting effect is evident in the polarization maps (fifth row), where the presence of elliptical polarization confirms this interaction.

Finally, a comparison between the numerical predictions in Fig.\ref{fig:resultModel} and the experimental measurements in Fig.\ref{fig:resultExp} reveals a good agreement between theory and experiment (up to an inessential global phase factor, which is responsible for the different orientation of the experimentally measured and theoretically calculated mode structures in Figs. \ref{fig:resultExp} and \ref{fig:resultModel}). This agreement validates the accuracy of theoretical framework and confirms that the sT-RCFs support robust amplification of OAM1 and OAM2 modes, maintaining high modal purity.

\section{Conclusion}
In summary, we demonstrated the Yb-doped ring-shaped core spun tapered double-clad fiber as a robust solution for efficient amplification of circularly-polarized $\ell=1$ and $\ell=2$ OAM modes preserving their coupled modal purity. The sT-RCF featured twisting geometry, which maintained circular intrinsic birefringence and supported propagation of circularly-polarized beams with minimal distortion. The ring-shaped core enabled efficient amplification of doughnut-shaped beams in the desired mode with conservation of their coupled power fraction. The amplification of OAM1 carrying 60 ps at 15 MHz repetition rate at 1030 nm resulted in 1.3 W output power with a gain level of 11.87 dB. The coupled OAM1 mode fraction of 95\% remained unchanged during the amplification, while the residual power was distributed between right-handed OAM 1 and fundamental mode. For OAM2, the output reached 1.25 W with 12 dB amplification and 97\% of the mode purity. For both modes, the circularly polarized topology remained largely unchanged with a small deviation in ellipticity due to the weak coupling into neighbor modes. The experimental observations were in good agreement with the theoretical modeling, and provided a consistent description of OAM mode evolution in the twisted, tapered active ring-core geometry. The preservation of mode purity and polarization topology during the amplification in sT-RCF demonstrated a promising route towards power scaling of OAM beams of exceptionally high quality.

%\author{Author One\authormark{1} and Author Two\authormark{2,*}}

%\address{\authormark{1}Peer Review, Publications Department,
%Optica Publishing Group, 2010 Massachusetts Avenue NW,
%Washington, DC 20036, USA\\
%\authormark{2}Publications Department, Optica Publishing Group,
%2010 Massachusetts Avenue NW, Washington, DC 20036, USA\\
%%\authormark{3}xyz@optica.org}

%\email{\authormark{*}xyz@optica.org}}

%Example with the corresponding author designated by an asterisk and a note indicating equal contributions by two authors.

%\author{Author One\authormark{1,3} and Author %Two\authormark{2,3,*}}

%\address{\authormark{1}Peer Review, Publications Department,
%Optica Publishing Group, 2010 Massachusetts Avenue NW, %Washington, DC 20036, USA\\
%\authormark{2}Publications Department, Optica Publishing Group, %2010 Massachusetts Avenue NW, Washington, DC 20036, USA\\
%\authormark{3}The authors contributed equally to this work.\\
%\authormark{*}xyz@optica.org}}

%\section{Examples of Article Components}
%\label{sec:examples}

%\l\section{Back matter}
%\lBack matter sections should be listed in the order Funding/Acknowledgement/Disclosures/Data Availability %\lStatement/Supplemental Document section. 

\section{backmatter}

\subsection{Funding} This work has been supported by the V4F project (grant agreement number 101096317), funded by the European Commission Horizon Europe Pathfinder-OPEN Program, the Research Council of Finland Flagship Programme, Photonics Research and Innovation (PREIN) (decision 320165) and Actphast 4.0 EU Horizon 2020 research and innovation program under GA No 779472 ( project P2020-34).

\subsection{Acknowledgment} K.W. would like to thank Michael Teupser (Leibniz IPHT) for RIP measurements and Jan Dellith (Leibniz IPHT) for EPMA analysis of core preform samples.

\subsection{Disclosures} The authors declare no conflicts of interest.

\bigskip

% Bibliography
%\bibliography{apssamp}

\begin{thebibliography}{43}%
\makeatletter
\providecommand \@ifxundefined [1]{%
 \@ifx{#1\undefined}
}%
\providecommand \@ifnum [1]{%
 \ifnum #1\expandafter \@firstoftwo
 \else \expandafter \@secondoftwo
 \fi
}%
\providecommand \@ifx [1]{%
 \ifx #1\expandafter \@firstoftwo
 \else \expandafter \@secondoftwo
 \fi
}%
\providecommand \natexlab [1]{#1}%
\providecommand \enquote  [1]{``#1''}%
\providecommand \bibnamefont  [1]{#1}%
\providecommand \bibfnamefont [1]{#1}%
\providecommand \citenamefont [1]{#1}%
\providecommand \href@noop [0]{\@secondoftwo}%
\providecommand \href [0]{\begingroup \@sanitize@url \@href}%
\providecommand \@href[1]{\@@startlink{#1}\@@href}%
\providecommand \@@href[1]{\endgroup#1\@@endlink}%
\providecommand \@sanitize@url [0]{\catcode `\\12\catcode `\$12\catcode `\&12\catcode `\#12\catcode `\^12\catcode `\_12\catcode `\%12\relax}%
\providecommand \@@startlink[1]{}%
\providecommand \@@endlink[0]{}%
\providecommand \url  [0]{\begingroup\@sanitize@url \@url }%
\providecommand \@url [1]{\endgroup\@href {#1}{\urlprefix }}%
\providecommand \urlprefix  [0]{URL }%
\providecommand \Eprint [0]{\href }%
\providecommand \doibase [0]{https://doi.org/}%
\providecommand \selectlanguage [0]{\@gobble}%
\providecommand \bibinfo  [0]{\@secondoftwo}%
\providecommand \bibfield  [0]{\@secondoftwo}%
\providecommand \translation [1]{[#1]}%
\providecommand \BibitemOpen [0]{}%
\providecommand \bibitemStop [0]{}%
\providecommand \bibitemNoStop [0]{.\EOS\space}%
\providecommand \EOS [0]{\spacefactor3000\relax}%
\providecommand \BibitemShut  [1]{\csname bibitem#1\endcsname}%
\let\auto@bib@innerbib\@empty
%</preamble>
\bibitem [{\citenamefont {Andrew}\ and\ \citenamefont {Babiker}(2013)}]{OAMlight}%
  \BibitemOpen
  \bibinfo {editor} {\bibfnamefont {D.}~\bibnamefont {Andrew}}\ and\ \bibinfo {editor} {\bibfnamefont {M.}~\bibnamefont {Babiker}},\ eds.,\ \href@noop {} {\emph {\bibinfo {title} {The Angular Momentum of Light}}}\ (\bibinfo  {publisher} {Cambridge University Press},\ \bibinfo {year} {2013})\BibitemShut {NoStop}%
\bibitem [{\citenamefont {Nye}\ and\ \citenamefont {Berry}(1974)}]{berryNye}%
  \BibitemOpen
  \bibfield  {author} {\bibinfo {author} {\bibfnamefont {J.~F.}\ \bibnamefont {Nye}}\ and\ \bibinfo {author} {\bibfnamefont {M.~V.}\ \bibnamefont {Berry}},\ }\bibfield  {title} {\bibinfo {title} {Dislocations in wave trains},\ }\href {https://doi.org/10.1098/rspa.1974.0012} {\bibfield  {journal} {\bibinfo  {journal} {Proceedings of the Royal Society of London. A. Mathematical and Physical Sciences}\ }\textbf {\bibinfo {volume} {336}},\ \bibinfo {pages} {165} (\bibinfo {year} {1974})},\ \Eprint {https://arxiv.org/abs/https://royalsocietypublishing.org/rspa/article-pdf/336/1605/165/60607/rspa.1974.0012.pdf} {https://royalsocietypublishing.org/rspa/article-pdf/336/1605/165/60607/rspa.1974.0012.pdf} \BibitemShut {NoStop}%
\bibitem [{\citenamefont {Allen}\ \emph {et~al.}(1992)\citenamefont {Allen}, \citenamefont {Beijersbergen}, \citenamefont {Spreeuw},\ and\ \citenamefont {Woerdman}}]{woerdman}%
  \BibitemOpen
  \bibfield  {author} {\bibinfo {author} {\bibfnamefont {L.}~\bibnamefont {Allen}}, \bibinfo {author} {\bibfnamefont {M.~W.}\ \bibnamefont {Beijersbergen}}, \bibinfo {author} {\bibfnamefont {R.~J.~C.}\ \bibnamefont {Spreeuw}},\ and\ \bibinfo {author} {\bibfnamefont {J.~P.}\ \bibnamefont {Woerdman}},\ }\bibfield  {title} {\bibinfo {title} {Orbital angular momentum of light and the transformation of \uppercase{L}aguerre-\uppercase{G}aussian laser modes},\ }\href {https://doi.org/10.1103/PhysRevA.45.8185} {\bibfield  {journal} {\bibinfo  {journal} {Phys. Rev. A}\ }\textbf {\bibinfo {volume} {45}},\ \bibinfo {pages} {8185} (\bibinfo {year} {1992})}\BibitemShut {NoStop}%
\bibitem [{\citenamefont {Singh}\ \emph {et~al.}(2017)\citenamefont {Singh}, \citenamefont {Nagar}, \citenamefont {Roichman},\ and\ \citenamefont {Arie}}]{partManipulation}%
  \BibitemOpen
  \bibfield  {author} {\bibinfo {author} {\bibfnamefont {B.}~\bibnamefont {Singh}}, \bibinfo {author} {\bibfnamefont {H.}~\bibnamefont {Nagar}}, \bibinfo {author} {\bibfnamefont {Y.}~\bibnamefont {Roichman}},\ and\ \bibinfo {author} {\bibfnamefont {A.}~\bibnamefont {Arie}},\ }\bibfield  {title} {\bibinfo {title} {Particle manipulation beyond the diffraction limit using structured super-oscillating light beams},\ }\href@noop {} {\bibfield  {journal} {\bibinfo  {journal} {Light Science and Applications}\ }\textbf {\bibinfo {volume} {6}},\ \bibinfo {pages} {17050} (\bibinfo {year} {2017})}\BibitemShut {NoStop}%
\bibitem [{\citenamefont {Hnatovsky}\ \emph {et~al.}(2010)\citenamefont {Hnatovsky}, \citenamefont {Shvedov}, \citenamefont {Krolikowski},\ and\ \citenamefont {Rode}}]{Hnatovsky:10}%
  \BibitemOpen
  \bibfield  {author} {\bibinfo {author} {\bibfnamefont {C.}~\bibnamefont {Hnatovsky}}, \bibinfo {author} {\bibfnamefont {V.~G.}\ \bibnamefont {Shvedov}}, \bibinfo {author} {\bibfnamefont {W.}~\bibnamefont {Krolikowski}},\ and\ \bibinfo {author} {\bibfnamefont {A.~V.}\ \bibnamefont {Rode}},\ }\bibfield  {title} {\bibinfo {title} {Materials processing with a tightly focused femtosecond laser vortex pulse},\ }\href {https://doi.org/10.1364/OL.35.003417} {\bibfield  {journal} {\bibinfo  {journal} {Opt. Lett.}\ }\textbf {\bibinfo {volume} {35}},\ \bibinfo {pages} {3417} (\bibinfo {year} {2010})}\BibitemShut {NoStop}%
\bibitem [{\citenamefont {Toyoda}\ \emph {et~al.}(2013)\citenamefont {Toyoda}, \citenamefont {Takahashi}, \citenamefont {Takizawa}, \citenamefont {Tokizane}, \citenamefont {Miyamoto}, \citenamefont {Morita},\ and\ \citenamefont {Omatsu}}]{PhysRevLett.110.143603}%
  \BibitemOpen
  \bibfield  {author} {\bibinfo {author} {\bibfnamefont {K.}~\bibnamefont {Toyoda}}, \bibinfo {author} {\bibfnamefont {F.}~\bibnamefont {Takahashi}}, \bibinfo {author} {\bibfnamefont {S.}~\bibnamefont {Takizawa}}, \bibinfo {author} {\bibfnamefont {Y.}~\bibnamefont {Tokizane}}, \bibinfo {author} {\bibfnamefont {K.}~\bibnamefont {Miyamoto}}, \bibinfo {author} {\bibfnamefont {R.}~\bibnamefont {Morita}},\ and\ \bibinfo {author} {\bibfnamefont {T.}~\bibnamefont {Omatsu}},\ }\bibfield  {title} {\bibinfo {title} {Transfer of light helicity to nanostructures},\ }\href {https://doi.org/10.1103/PhysRevLett.110.143603} {\bibfield  {journal} {\bibinfo  {journal} {Phys. Rev. Lett.}\ }\textbf {\bibinfo {volume} {110}},\ \bibinfo {pages} {143603} (\bibinfo {year} {2013})}\BibitemShut {NoStop}%
\bibitem [{\citenamefont {Prajapati}\ \emph {et~al.}(2019)\citenamefont {Prajapati}, \citenamefont {Super}, \citenamefont {Lanning}, \citenamefont {Dowling},\ and\ \citenamefont {Novikova}}]{nonlinearFWM1}%
  \BibitemOpen
  \bibfield  {author} {\bibinfo {author} {\bibfnamefont {N.}~\bibnamefont {Prajapati}}, \bibinfo {author} {\bibfnamefont {N.}~\bibnamefont {Super}}, \bibinfo {author} {\bibfnamefont {N.~R.}\ \bibnamefont {Lanning}}, \bibinfo {author} {\bibfnamefont {J.~P.}\ \bibnamefont {Dowling}},\ and\ \bibinfo {author} {\bibfnamefont {I.}~\bibnamefont {Novikova}},\ }\bibfield  {title} {\bibinfo {title} {Optical angular momentum manipulations in a four-wave mixing process},\ }\href {https://doi.org/10.1364/OL.44.000739} {\bibfield  {journal} {\bibinfo  {journal} {Opt. Lett.}\ }\textbf {\bibinfo {volume} {44}},\ \bibinfo {pages} {739} (\bibinfo {year} {2019})}\BibitemShut {NoStop}%
\bibitem [{\citenamefont {Buono}\ and\ \citenamefont {Forbes}(2022)}]{nonlinearFWM2}%
  \BibitemOpen
  \bibfield  {author} {\bibinfo {author} {\bibfnamefont {W.~T.}\ \bibnamefont {Buono}}\ and\ \bibinfo {author} {\bibfnamefont {A.}~\bibnamefont {Forbes}},\ }\bibfield  {title} {\bibinfo {title} {Nonlinear optics with structured light},\ }\href@noop {} {\bibfield  {journal} {\bibinfo  {journal} {Opto-Electron. Adv.}\ }\textbf {\bibinfo {volume} {5}},\ \bibinfo {pages} {210174} (\bibinfo {year} {2022})}\BibitemShut {NoStop}%
\bibitem [{\citenamefont {da~Motta}\ and\ \citenamefont {Vianna}(2025)}]{nonlinearFWM3}%
  \BibitemOpen
  \bibfield  {author} {\bibinfo {author} {\bibfnamefont {M.~R.~L.}\ \bibnamefont {da~Motta}}\ and\ \bibinfo {author} {\bibfnamefont {S.~S.}\ \bibnamefont {Vianna}},\ }\bibfield  {title} {\bibinfo {title} {Spatial correlations in four-wave mixing with structured light},\ }\href {https://doi.org/10.1103/33sn-41cs} {\bibfield  {journal} {\bibinfo  {journal} {Phys. Rev. A}\ }\textbf {\bibinfo {volume} {112}},\ \bibinfo {pages} {063714} (\bibinfo {year} {2025})}\BibitemShut {NoStop}%
\bibitem [{\citenamefont {da~Motta}\ \emph {et~al.}(2024)\citenamefont {da~Motta}, \citenamefont {Alves}, \citenamefont {Khoury},\ and\ \citenamefont {Vianna}}]{nonlinearFWM4}%
  \BibitemOpen
  \bibfield  {author} {\bibinfo {author} {\bibfnamefont {M.~R.~L.}\ \bibnamefont {da~Motta}}, \bibinfo {author} {\bibfnamefont {G.~B.}\ \bibnamefont {Alves}}, \bibinfo {author} {\bibfnamefont {A.~Z.}\ \bibnamefont {Khoury}},\ and\ \bibinfo {author} {\bibfnamefont {S.~S.}\ \bibnamefont {Vianna}},\ }\bibfield  {title} {\bibinfo {title} {Poincar\'e-sphere symmetries in four-wave mixing with orbital angular momentum},\ }\href {https://doi.org/10.1103/PhysRevA.109.013506} {\bibfield  {journal} {\bibinfo  {journal} {Phys. Rev. A}\ }\textbf {\bibinfo {volume} {109}},\ \bibinfo {pages} {013506} (\bibinfo {year} {2024})}\BibitemShut {NoStop}%
\bibitem [{\citenamefont {Al-Amri}\ \emph {et~al.}(2013)\citenamefont {Al-Amri}, \citenamefont {Andrews},\ and\ \citenamefont {Babiker}}]{OAMcomm}%
  \BibitemOpen
  \bibinfo {editor} {\bibfnamefont {M.~D.}\ \bibnamefont {Al-Amri}}, \bibinfo {editor} {\bibfnamefont {D.}~\bibnamefont {Andrews}},\ and\ \bibinfo {editor} {\bibfnamefont {M.}~\bibnamefont {Babiker}},\ eds.,\ \href@noop {} {\emph {\bibinfo {title} {Structured Light for Optical Communication}}},\ \bibinfo {edition} {first edition}\ ed.\ (\bibinfo  {publisher} {Elsevier},\ \bibinfo {year} {2013})\BibitemShut {NoStop}%
\bibitem [{\citenamefont {Chen}\ \emph {et~al.}(2023)\citenamefont {Chen}, \citenamefont {Zhang}, \citenamefont {Han}, \citenamefont {Cao}, \citenamefont {Wand}, \citenamefont {Li}, \citenamefont {Doi}, \citenamefont {Kim},\ and\ \citenamefont {Xi}}]{superResImaging}%
  \BibitemOpen
  \bibfield  {author} {\bibinfo {author} {\bibfnamefont {X.}~\bibnamefont {Chen}}, \bibinfo {author} {\bibfnamefont {S.}~\bibnamefont {Zhang}}, \bibinfo {author} {\bibfnamefont {Y.}~\bibnamefont {Han}}, \bibinfo {author} {\bibfnamefont {R.}~\bibnamefont {Cao}}, \bibinfo {author} {\bibfnamefont {W.}~\bibnamefont {Wand}}, \bibinfo {author} {\bibfnamefont {D.}~\bibnamefont {Li}}, \bibinfo {author} {\bibfnamefont {Q.}~\bibnamefont {Doi}}, \bibinfo {author} {\bibfnamefont {D.}~\bibnamefont {Kim}},\ and\ \bibinfo {author} {\bibfnamefont {P.}~\bibnamefont {Xi}},\ }\bibfield  {title} {\bibinfo {title} {Superresolution structured illumination microscopy algorithms: a review},\ }\href@noop {} {\bibfield  {journal} {\bibinfo  {journal} {Light Science and Applications}\ }\textbf {\bibinfo {volume} {12}},\ \bibinfo {pages} {172} (\bibinfo {year} {2023})}\BibitemShut {NoStop}%
\bibitem [{\citenamefont {Baio}\ \emph {et~al.}(2020)\citenamefont {Baio}, \citenamefont {Robb}, \citenamefont {Yao},\ and\ \citenamefont {Oppo}}]{atomtronics}%
  \BibitemOpen
  \bibfield  {author} {\bibinfo {author} {\bibfnamefont {G.}~\bibnamefont {Baio}}, \bibinfo {author} {\bibfnamefont {G.~R.~M.}\ \bibnamefont {Robb}}, \bibinfo {author} {\bibfnamefont {A.~M.}\ \bibnamefont {Yao}},\ and\ \bibinfo {author} {\bibfnamefont {G.-L.}\ \bibnamefont {Oppo}},\ }\bibfield  {title} {\bibinfo {title} {Optomechanical transport of cold atoms induced by structured light},\ }\href {https://doi.org/10.1103/PhysRevResearch.2.023126} {\bibfield  {journal} {\bibinfo  {journal} {Phys. Rev. Res.}\ }\textbf {\bibinfo {volume} {2}},\ \bibinfo {pages} {023126} (\bibinfo {year} {2020})}\BibitemShut {NoStop}%
\bibitem [{\citenamefont {Bozinovic}\ \emph {et~al.}(2013)\citenamefont {Bozinovic}, \citenamefont {Yue}, \citenamefont {Ren}, \citenamefont {Tur}, \citenamefont {Kristensen}, \citenamefont {Huang}, \citenamefont {Willner},\ and\ \citenamefont {Ramachandran}}]{Bozinovic2013}%
  \BibitemOpen
  \bibfield  {author} {\bibinfo {author} {\bibfnamefont {N.}~\bibnamefont {Bozinovic}}, \bibinfo {author} {\bibfnamefont {Y.}~\bibnamefont {Yue}}, \bibinfo {author} {\bibfnamefont {Y.}~\bibnamefont {Ren}}, \bibinfo {author} {\bibfnamefont {M.}~\bibnamefont {Tur}}, \bibinfo {author} {\bibfnamefont {P.}~\bibnamefont {Kristensen}}, \bibinfo {author} {\bibfnamefont {H.}~\bibnamefont {Huang}}, \bibinfo {author} {\bibfnamefont {A.~E.}\ \bibnamefont {Willner}},\ and\ \bibinfo {author} {\bibfnamefont {S.}~\bibnamefont {Ramachandran}},\ }\bibfield  {title} {\bibinfo {title} {Terabit-scale orbital angular momentum mode division multiplexing in fibers},\ }\href {https://doi.org/10.1126/science.1237861} {\bibfield  {journal} {\bibinfo  {journal} {Science}\ }\textbf {\bibinfo {volume} {340}},\ \bibinfo {pages} {1545} (\bibinfo {year} {2013})},\ \Eprint {https://arxiv.org/abs/https://www.science.org/doi/pdf/10.1126/science.1237861} {https://www.science.org/doi/pdf/10.1126/science.1237861} \BibitemShut {NoStop}%
\bibitem [{\citenamefont {Erhard}\ \emph {et~al.}(2018)\citenamefont {Erhard}, \citenamefont {Fickler}, \citenamefont {Krenn},\ and\ \citenamefont {Zeilinger}}]{quantOptics}%
  \BibitemOpen
  \bibfield  {author} {\bibinfo {author} {\bibfnamefont {M.}~\bibnamefont {Erhard}}, \bibinfo {author} {\bibfnamefont {R.}~\bibnamefont {Fickler}}, \bibinfo {author} {\bibfnamefont {M.}~\bibnamefont {Krenn}},\ and\ \bibinfo {author} {\bibfnamefont {A.}~\bibnamefont {Zeilinger}},\ }\bibfield  {title} {\bibinfo {title} {Twisted photons: new quantum perspectives in high dimensions},\ }\href@noop {} {\bibfield  {journal} {\bibinfo  {journal} {Light Science and Applications}\ }\textbf {\bibinfo {volume} {7}},\ \bibinfo {pages} {17146} (\bibinfo {year} {2018})}\BibitemShut {NoStop}%
\bibitem [{\citenamefont {Rubinsztein-Dunlop}\ \emph {et~al.}(2016)\citenamefont {Rubinsztein-Dunlop}, \citenamefont {Forbes}, \citenamefont {Berry}, \citenamefont {Dennis}, \citenamefont {Andrews}, \citenamefont {Mansuripur}, \citenamefont {Denz}, \citenamefont {Alpmann}, \citenamefont {Banzer}, \citenamefont {Bauer}, \citenamefont {Karimi}, \citenamefont {Marrucci}, \citenamefont {Padgett}, \citenamefont {Ritsch-Marte}, \citenamefont {Litchinitser}, \citenamefont {Bigelow}, \citenamefont {Rosales-Guzmán}, \citenamefont {Belmonte}, \citenamefont {Torres}, \citenamefont {Neely}, \citenamefont {Baker}, \citenamefont {Gordon}, \citenamefont {Stilgoe}, \citenamefont {Romero}, \citenamefont {White}, \citenamefont {Fickler}, \citenamefont {Willner}, \citenamefont {Xie}, \citenamefont {McMorran},\ and\ \citenamefont {Weiner}}]{OAMroadmap}%
  \BibitemOpen
  \bibfield  {author} {\bibinfo {author} {\bibfnamefont {H.}~\bibnamefont {Rubinsztein-Dunlop}}, \bibinfo {author} {\bibfnamefont {A.}~\bibnamefont {Forbes}}, \bibinfo {author} {\bibfnamefont {M.~V.}\ \bibnamefont {Berry}}, \bibinfo {author} {\bibfnamefont {M.~R.}\ \bibnamefont {Dennis}}, \bibinfo {author} {\bibfnamefont {D.~L.}\ \bibnamefont {Andrews}}, \bibinfo {author} {\bibfnamefont {M.}~\bibnamefont {Mansuripur}}, \bibinfo {author} {\bibfnamefont {C.}~\bibnamefont {Denz}}, \bibinfo {author} {\bibfnamefont {C.}~\bibnamefont {Alpmann}}, \bibinfo {author} {\bibfnamefont {P.}~\bibnamefont {Banzer}}, \bibinfo {author} {\bibfnamefont {T.}~\bibnamefont {Bauer}}, \bibinfo {author} {\bibfnamefont {E.}~\bibnamefont {Karimi}}, \bibinfo {author} {\bibfnamefont {L.}~\bibnamefont {Marrucci}}, \bibinfo {author} {\bibfnamefont {M.}~\bibnamefont {Padgett}}, \bibinfo {author} {\bibfnamefont {M.}~\bibnamefont {Ritsch-Marte}}, \bibinfo {author} {\bibfnamefont {N.~M.}\ \bibnamefont {Litchinitser}}, \bibinfo {author}
  {\bibfnamefont {N.~P.}\ \bibnamefont {Bigelow}}, \bibinfo {author} {\bibfnamefont {C.}~\bibnamefont {Rosales-Guzmán}}, \bibinfo {author} {\bibfnamefont {A.}~\bibnamefont {Belmonte}}, \bibinfo {author} {\bibfnamefont {J.~P.}\ \bibnamefont {Torres}}, \bibinfo {author} {\bibfnamefont {T.~W.}\ \bibnamefont {Neely}}, \bibinfo {author} {\bibfnamefont {M.}~\bibnamefont {Baker}}, \bibinfo {author} {\bibfnamefont {R.}~\bibnamefont {Gordon}}, \bibinfo {author} {\bibfnamefont {A.~B.}\ \bibnamefont {Stilgoe}}, \bibinfo {author} {\bibfnamefont {J.}~\bibnamefont {Romero}}, \bibinfo {author} {\bibfnamefont {A.~G.}\ \bibnamefont {White}}, \bibinfo {author} {\bibfnamefont {R.}~\bibnamefont {Fickler}}, \bibinfo {author} {\bibfnamefont {A.~E.}\ \bibnamefont {Willner}}, \bibinfo {author} {\bibfnamefont {G.}~\bibnamefont {Xie}}, \bibinfo {author} {\bibfnamefont {B.}~\bibnamefont {McMorran}},\ and\ \bibinfo {author} {\bibfnamefont {A.~M.}\ \bibnamefont {Weiner}},\ }\bibfield  {title} {\bibinfo {title} {Roadmap on structured
  light},\ }\href {https://doi.org/10.1088/2040-8978/19/1/013001} {\bibfield  {journal} {\bibinfo  {journal} {Journal of Optics}\ }\textbf {\bibinfo {volume} {19}},\ \bibinfo {pages} {013001} (\bibinfo {year} {2016})}\BibitemShut {NoStop}%
\bibitem [{\citenamefont {Zhu}\ and\ \citenamefont {Wang}(2014)}]{SLMgeneration}%
  \BibitemOpen
  \bibfield  {author} {\bibinfo {author} {\bibfnamefont {L.}~\bibnamefont {Zhu}}\ and\ \bibinfo {author} {\bibfnamefont {J.}~\bibnamefont {Wang}},\ }\bibfield  {title} {\bibinfo {title} {Arbitrary manipulation of spatial amplitude and phase using phase-only spatial light modulators},\ }\href@noop {} {\bibfield  {journal} {\bibinfo  {journal} {Sci-Rep.}\ }\textbf {\bibinfo {volume} {4}},\ \bibinfo {pages} {7441} (\bibinfo {year} {2014})}\BibitemShut {NoStop}%
\bibitem [{\citenamefont {Marrucci}\ \emph {et~al.}(2006)\citenamefont {Marrucci}, \citenamefont {Manzo},\ and\ \citenamefont {Paparo}}]{qPlate}%
  \BibitemOpen
  \bibfield  {author} {\bibinfo {author} {\bibfnamefont {L.}~\bibnamefont {Marrucci}}, \bibinfo {author} {\bibfnamefont {C.}~\bibnamefont {Manzo}},\ and\ \bibinfo {author} {\bibfnamefont {D.}~\bibnamefont {Paparo}},\ }\bibfield  {title} {\bibinfo {title} {Optical spin-to-orbital angular momentum conversion in inhomogeneous anisotropic media},\ }\href {https://doi.org/10.1103/PhysRevLett.96.163905} {\bibfield  {journal} {\bibinfo  {journal} {Phys. Rev. Lett.}\ }\textbf {\bibinfo {volume} {96}},\ \bibinfo {pages} {163905} (\bibinfo {year} {2006})}\BibitemShut {NoStop}%
\bibitem [{\citenamefont {Jain}\ \emph {et~al.}(2015)\citenamefont {Jain}, \citenamefont {Jung}, \citenamefont {Barua}, \citenamefont {Alam},\ and\ \citenamefont {Sahu}}]{Jain:15}%
  \BibitemOpen
  \bibfield  {author} {\bibinfo {author} {\bibfnamefont {D.}~\bibnamefont {Jain}}, \bibinfo {author} {\bibfnamefont {Y.}~\bibnamefont {Jung}}, \bibinfo {author} {\bibfnamefont {P.}~\bibnamefont {Barua}}, \bibinfo {author} {\bibfnamefont {S.}~\bibnamefont {Alam}},\ and\ \bibinfo {author} {\bibfnamefont {J.~K.}\ \bibnamefont {Sahu}},\ }\bibfield  {title} {\bibinfo {title} {Demonstration of ultra-low \uppercase{NA} rare-earth doped step index fiber for applications in high power fiber lasers},\ }\href {https://doi.org/10.1364/OE.23.007407} {\bibfield  {journal} {\bibinfo  {journal} {Opt. Express}\ }\textbf {\bibinfo {volume} {23}},\ \bibinfo {pages} {7407} (\bibinfo {year} {2015})}\BibitemShut {NoStop}%
\bibitem [{\citenamefont {Limpert}\ \emph {et~al.}(2007)\citenamefont {Limpert}, \citenamefont {Roser}, \citenamefont {Klingebiel}, \citenamefont {Schreiber}, \citenamefont {Wirth}, \citenamefont {Peschel}, \citenamefont {Eberhardt},\ and\ \citenamefont {Tunnermann}}]{Limpert2007}%
  \BibitemOpen
  \bibfield  {author} {\bibinfo {author} {\bibfnamefont {J.}~\bibnamefont {Limpert}}, \bibinfo {author} {\bibfnamefont {F.}~\bibnamefont {Roser}}, \bibinfo {author} {\bibfnamefont {S.}~\bibnamefont {Klingebiel}}, \bibinfo {author} {\bibfnamefont {T.}~\bibnamefont {Schreiber}}, \bibinfo {author} {\bibfnamefont {C.}~\bibnamefont {Wirth}}, \bibinfo {author} {\bibfnamefont {T.}~\bibnamefont {Peschel}}, \bibinfo {author} {\bibfnamefont {R.}~\bibnamefont {Eberhardt}},\ and\ \bibinfo {author} {\bibfnamefont {A.}~\bibnamefont {Tunnermann}},\ }\bibfield  {title} {\bibinfo {title} {The rising power of fiber lasers and amplifiers},\ }\href {https://doi.org/10.1109/JSTQE.2007.897182} {\bibfield  {journal} {\bibinfo  {journal} {IEEE Journal of Selected Topics in Quantum Electronics}\ }\textbf {\bibinfo {volume} {13}},\ \bibinfo {pages} {537} (\bibinfo {year} {2007})}\BibitemShut {NoStop}%
\bibitem [{\citenamefont {Fathi}\ \emph {et~al.}(2025)\citenamefont {Fathi}, \citenamefont {Aghayari}, \citenamefont {Grishchenko}, \citenamefont {Yadav}, \citenamefont {Rafailov}, \citenamefont {Gumenyuk},\ and\ \citenamefont {Filippov}}]{fathi2025versatile}%
  \BibitemOpen
  \bibfield  {author} {\bibinfo {author} {\bibfnamefont {H.}~\bibnamefont {Fathi}}, \bibinfo {author} {\bibfnamefont {E.}~\bibnamefont {Aghayari}}, \bibinfo {author} {\bibfnamefont {A.}~\bibnamefont {Grishchenko}}, \bibinfo {author} {\bibfnamefont {A.}~\bibnamefont {Yadav}}, \bibinfo {author} {\bibfnamefont {E.}~\bibnamefont {Rafailov}}, \bibinfo {author} {\bibfnamefont {R.}~\bibnamefont {Gumenyuk}},\ and\ \bibinfo {author} {\bibfnamefont {V.}~\bibnamefont {Filippov}},\ }\bibfield  {title} {\bibinfo {title} {Versatile high-power monolithic all-glass fiber amplifier for pulsed signals with a wide range of repetition rates},\ }\href@noop {} {\bibfield  {journal} {\bibinfo  {journal} {Scientific Reports}\ } (\bibinfo {year} {2025})}\BibitemShut {NoStop}%
\bibitem [{\citenamefont {Sun}\ \emph {et~al.}(2022)\citenamefont {Sun}, \citenamefont {Liu}, \citenamefont {Han}, \citenamefont {Zhu}, \citenamefont {Shen},\ and\ \citenamefont {Yang}}]{Sun:22}%
  \BibitemOpen
  \bibfield  {author} {\bibinfo {author} {\bibfnamefont {J.}~\bibnamefont {Sun}}, \bibinfo {author} {\bibfnamefont {L.}~\bibnamefont {Liu}}, \bibinfo {author} {\bibfnamefont {L.}~\bibnamefont {Han}}, \bibinfo {author} {\bibfnamefont {Q.}~\bibnamefont {Zhu}}, \bibinfo {author} {\bibfnamefont {X.}~\bibnamefont {Shen}},\ and\ \bibinfo {author} {\bibfnamefont {K.}~\bibnamefont {Yang}},\ }\bibfield  {title} {\bibinfo {title} {100 k\uppercase{W} ultra high power fiber laser},\ }\href {https://doi.org/10.1364/OPTCON.465836} {\bibfield  {journal} {\bibinfo  {journal} {Opt. Continuum}\ }\textbf {\bibinfo {volume} {1}},\ \bibinfo {pages} {1932} (\bibinfo {year} {2022})}\BibitemShut {NoStop}%
\bibitem [{\citenamefont {Lin}\ \emph {et~al.}(2017)\citenamefont {Lin}, \citenamefont {Baktash}, \citenamefont {Berendt}, \citenamefont {Beresna}, \citenamefont {Kazansky}, \citenamefont {Clarkson}, \citenamefont {Alam},\ and\ \citenamefont {Richardson}}]{Lin:17}%
  \BibitemOpen
  \bibfield  {author} {\bibinfo {author} {\bibfnamefont {D.}~\bibnamefont {Lin}}, \bibinfo {author} {\bibfnamefont {N.}~\bibnamefont {Baktash}}, \bibinfo {author} {\bibfnamefont {M.}~\bibnamefont {Berendt}}, \bibinfo {author} {\bibfnamefont {M.}~\bibnamefont {Beresna}}, \bibinfo {author} {\bibfnamefont {P.~G.}\ \bibnamefont {Kazansky}}, \bibinfo {author} {\bibfnamefont {W.~A.}\ \bibnamefont {Clarkson}}, \bibinfo {author} {\bibfnamefont {S.~U.}\ \bibnamefont {Alam}},\ and\ \bibinfo {author} {\bibfnamefont {D.~J.}\ \bibnamefont {Richardson}},\ }\bibfield  {title} {\bibinfo {title} {Radially and azimuthally polarized nanosecond \uppercase{Y}b-doped fiber \uppercase{MOPA} system incorporating temporal shaping},\ }\href {https://doi.org/10.1364/OL.42.001740} {\bibfield  {journal} {\bibinfo  {journal} {Opt. Lett.}\ }\textbf {\bibinfo {volume} {42}},\ \bibinfo {pages} {1740} (\bibinfo {year} {2017})}\BibitemShut {NoStop}%
\bibitem [{\citenamefont {Liu}\ \emph {et~al.}(2024)\citenamefont {Liu}, \citenamefont {Shi}, \citenamefont {Liu}, \citenamefont {Peng}, \citenamefont {Feng}, \citenamefont {Sun}, \citenamefont {Ma}, \citenamefont {Zhao}, \citenamefont {Gao}, \citenamefont {Liu},\ and\ \citenamefont {Tang}}]{Liu:24}%
  \BibitemOpen
  \bibfield  {author} {\bibinfo {author} {\bibfnamefont {S.}~\bibnamefont {Liu}}, \bibinfo {author} {\bibfnamefont {X.}~\bibnamefont {Shi}}, \bibinfo {author} {\bibfnamefont {H.}~\bibnamefont {Liu}}, \bibinfo {author} {\bibfnamefont {W.}~\bibnamefont {Peng}}, \bibinfo {author} {\bibfnamefont {Y.}~\bibnamefont {Feng}}, \bibinfo {author} {\bibfnamefont {Y.}~\bibnamefont {Sun}}, \bibinfo {author} {\bibfnamefont {Y.}~\bibnamefont {Ma}}, \bibinfo {author} {\bibfnamefont {Z.}~\bibnamefont {Zhao}}, \bibinfo {author} {\bibfnamefont {Q.}~\bibnamefont {Gao}}, \bibinfo {author} {\bibfnamefont {Z.}~\bibnamefont {Liu}},\ and\ \bibinfo {author} {\bibfnamefont {C.}~\bibnamefont {Tang}},\ }\bibfield  {title} {\bibinfo {title} {All-fiber hectowatt radially polarized laser system},\ }\href {https://doi.org/10.1364/OL.502874} {\bibfield  {journal} {\bibinfo  {journal} {Opt. Lett.}\ }\textbf {\bibinfo {volume} {49}},\ \bibinfo {pages} {891} (\bibinfo {year} {2024})}\BibitemShut {NoStop}%
\bibitem [{\citenamefont {Koyama}\ \emph {et~al.}(2011{\natexlab{a}})\citenamefont {Koyama}, \citenamefont {Hirose}, \citenamefont {Okida}, \citenamefont {Miyamoto},\ and\ \citenamefont {Omatsu}}]{Koyama:11pico}%
  \BibitemOpen
  \bibfield  {author} {\bibinfo {author} {\bibfnamefont {M.}~\bibnamefont {Koyama}}, \bibinfo {author} {\bibfnamefont {T.}~\bibnamefont {Hirose}}, \bibinfo {author} {\bibfnamefont {M.}~\bibnamefont {Okida}}, \bibinfo {author} {\bibfnamefont {K.}~\bibnamefont {Miyamoto}},\ and\ \bibinfo {author} {\bibfnamefont {T.}~\bibnamefont {Omatsu}},\ }\bibfield  {title} {\bibinfo {title} {Power scaling of a picosecond vortex laser based on a stressed \uppercase{Y}b-doped fiber amplifier},\ }\href {https://doi.org/10.1364/OE.19.000994} {\bibfield  {journal} {\bibinfo  {journal} {Opt. Express}\ }\textbf {\bibinfo {volume} {19}},\ \bibinfo {pages} {994} (\bibinfo {year} {2011}{\natexlab{a}})}\BibitemShut {NoStop}%
\bibitem [{\citenamefont {Ramachandran}\ and\ \citenamefont {Kristensen}(2013)}]{RamachandranKristensen2013}%
  \BibitemOpen
  \bibfield  {author} {\bibinfo {author} {\bibfnamefont {S.}~\bibnamefont {Ramachandran}}\ and\ \bibinfo {author} {\bibfnamefont {P.}~\bibnamefont {Kristensen}},\ }\bibfield  {title} {\bibinfo {title} {Optical vortices in fiber},\ }\href {https://doi.org/doi:10.1515/nanoph-2013-0047} {\bibfield  {journal} {\bibinfo  {journal} {Nanophotonics}\ }\textbf {\bibinfo {volume} {2}},\ \bibinfo {pages} {455} (\bibinfo {year} {2013})}\BibitemShut {NoStop}%
\bibitem [{\citenamefont {Koyama}\ \emph {et~al.}(2011{\natexlab{b}})\citenamefont {Koyama}, \citenamefont {Hirose}, \citenamefont {Okida}, \citenamefont {Miyamoto},\ and\ \citenamefont {Omatsu}}]{Koyama:11nano}%
  \BibitemOpen
  \bibfield  {author} {\bibinfo {author} {\bibfnamefont {M.}~\bibnamefont {Koyama}}, \bibinfo {author} {\bibfnamefont {T.}~\bibnamefont {Hirose}}, \bibinfo {author} {\bibfnamefont {M.}~\bibnamefont {Okida}}, \bibinfo {author} {\bibfnamefont {K.}~\bibnamefont {Miyamoto}},\ and\ \bibinfo {author} {\bibfnamefont {T.}~\bibnamefont {Omatsu}},\ }\bibfield  {title} {\bibinfo {title} {Nanosecond vortex laser pulses with millijoule pulse energies from a \uppercase{Y}b-doped double-clad fiber power amplifier},\ }\href {https://doi.org/10.1364/OE.19.014420} {\bibfield  {journal} {\bibinfo  {journal} {Opt. Express}\ }\textbf {\bibinfo {volume} {19}},\ \bibinfo {pages} {14420} (\bibinfo {year} {2011}{\natexlab{b}})}\BibitemShut {NoStop}%
\bibitem [{\citenamefont {Wu}\ \emph {et~al.}(2022)\citenamefont {Wu}, \citenamefont {Wen}, \citenamefont {Zhang}, \citenamefont {Cao}, \citenamefont {Chen}, \citenamefont {Zhang}, \citenamefont {Yusufu}, \citenamefont {Pang},\ and\ \citenamefont {Wang}}]{Wu:22}%
  \BibitemOpen
  \bibfield  {author} {\bibinfo {author} {\bibfnamefont {Y.}~\bibnamefont {Wu}}, \bibinfo {author} {\bibfnamefont {J.}~\bibnamefont {Wen}}, \bibinfo {author} {\bibfnamefont {M.}~\bibnamefont {Zhang}}, \bibinfo {author} {\bibfnamefont {Y.}~\bibnamefont {Cao}}, \bibinfo {author} {\bibfnamefont {W.}~\bibnamefont {Chen}}, \bibinfo {author} {\bibfnamefont {X.}~\bibnamefont {Zhang}}, \bibinfo {author} {\bibfnamefont {T.}~\bibnamefont {Yusufu}}, \bibinfo {author} {\bibfnamefont {F.}~\bibnamefont {Pang}},\ and\ \bibinfo {author} {\bibfnamefont {T.}~\bibnamefont {Wang}},\ }\bibfield  {title} {\bibinfo {title} {Low-loss and helical-phase-dependent selective excitation of high-order orbital angular momentum modes in a twisted ring-core fiber},\ }\href {https://doi.org/10.1364/OL.468259} {\bibfield  {journal} {\bibinfo  {journal} {Opt. Lett.}\ }\textbf {\bibinfo {volume} {47}},\ \bibinfo {pages} {4016} (\bibinfo {year} {2022})}\BibitemShut {NoStop}%
\bibitem [{\citenamefont {Russell}\ \emph {et~al.}(2017)\citenamefont {Russell}, \citenamefont {Beravat},\ and\ \citenamefont {Wong}}]{Russell2017}%
  \BibitemOpen
  \bibfield  {author} {\bibinfo {author} {\bibfnamefont {P.~S.}\ \bibnamefont {Russell}}, \bibinfo {author} {\bibfnamefont {R.}~\bibnamefont {Beravat}},\ and\ \bibinfo {author} {\bibfnamefont {G.~K.~L.}\ \bibnamefont {Wong}},\ }\bibfield  {title} {\bibinfo {title} {Helically twisted photonic crystal fibres},\ }\href {https://doi.org/10.1098/rsta.2015.0440} {\bibfield  {journal} {\bibinfo  {journal} {Philosophical Transactions of the Royal Society A: Mathematical, Physical and Engineering Sciences}\ }\textbf {\bibinfo {volume} {375}},\ \bibinfo {pages} {20150440} (\bibinfo {year} {2017})},\ \Eprint {https://arxiv.org/abs/https://royalsocietypublishing.org/rsta/article-pdf/doi/10.1098/rsta.2015.0440/1384890/rsta.2015.0440.pdf} {https://royalsocietypublishing.org/rsta/article-pdf/doi/10.1098/rsta.2015.0440/1384890/rsta.2015.0440.pdf} \BibitemShut {NoStop}%
\bibitem [{\citenamefont {Gregg}\ \emph {et~al.}(2015)\citenamefont {Gregg}, \citenamefont {Kristensen},\ and\ \citenamefont {Ramachandran}}]{Gregg:15}%
  \BibitemOpen
  \bibfield  {author} {\bibinfo {author} {\bibfnamefont {P.}~\bibnamefont {Gregg}}, \bibinfo {author} {\bibfnamefont {P.}~\bibnamefont {Kristensen}},\ and\ \bibinfo {author} {\bibfnamefont {S.}~\bibnamefont {Ramachandran}},\ }\bibfield  {title} {\bibinfo {title} {Conservation of orbital angular momentum in air-core optical fibers},\ }\href {https://doi.org/10.1364/OPTICA.2.000267} {\bibfield  {journal} {\bibinfo  {journal} {Optica}\ }\textbf {\bibinfo {volume} {2}},\ \bibinfo {pages} {267} (\bibinfo {year} {2015})}\BibitemShut {NoStop}%
\bibitem [{\citenamefont {Liu}\ \emph {et~al.}(2022)\citenamefont {Liu}, \citenamefont {Zhang}, \citenamefont {Liu}, \citenamefont {Lin}, \citenamefont {Li}, \citenamefont {Lin}, \citenamefont {Zhang}, \citenamefont {Huang}, \citenamefont {Mo}, \citenamefont {Shen}, \citenamefont {Lin}, \citenamefont {Chen}, \citenamefont {Gao}, \citenamefont {Zhang}, \citenamefont {Lan}, \citenamefont {Cai}, \citenamefont {Li},\ and\ \citenamefont {Yu}}]{Liu2022Nature}%
  \BibitemOpen
  \bibfield  {author} {\bibinfo {author} {\bibfnamefont {J.}~\bibnamefont {Liu}}, \bibinfo {author} {\bibfnamefont {J.}~\bibnamefont {Zhang}}, \bibinfo {author} {\bibfnamefont {J.}~\bibnamefont {Liu}}, \bibinfo {author} {\bibfnamefont {Z.}~\bibnamefont {Lin}}, \bibinfo {author} {\bibfnamefont {Z.}~\bibnamefont {Li}}, \bibinfo {author} {\bibfnamefont {Z.}~\bibnamefont {Lin}}, \bibinfo {author} {\bibfnamefont {J.}~\bibnamefont {Zhang}}, \bibinfo {author} {\bibfnamefont {C.}~\bibnamefont {Huang}}, \bibinfo {author} {\bibfnamefont {S.}~\bibnamefont {Mo}}, \bibinfo {author} {\bibfnamefont {L.}~\bibnamefont {Shen}}, \bibinfo {author} {\bibfnamefont {S.}~\bibnamefont {Lin}}, \bibinfo {author} {\bibfnamefont {Y.}~\bibnamefont {Chen}}, \bibinfo {author} {\bibfnamefont {R.}~\bibnamefont {Gao}}, \bibinfo {author} {\bibfnamefont {L.}~\bibnamefont {Zhang}}, \bibinfo {author} {\bibfnamefont {X.}~\bibnamefont {Lan}}, \bibinfo {author} {\bibfnamefont {X.}~\bibnamefont {Cai}}, \bibinfo {author} {\bibfnamefont
  {Z.}~\bibnamefont {Li}},\ and\ \bibinfo {author} {\bibfnamefont {S.}~\bibnamefont {Yu}},\ }\bibfield  {title} {\bibinfo {title} {1-\uppercase{P}bps orbital angular momentum fibre-optic transmission},\ }\href {https://doi.org/10.1038/s41377-022-00889-3} {\bibfield  {journal} {\bibinfo  {journal} {Light: Science \& Applications}\ }\textbf {\bibinfo {volume} {11}},\ \bibinfo {pages} {202} (\bibinfo {year} {2022})}\BibitemShut {NoStop}%
\bibitem [{\citenamefont {Wen}\ \emph {et~al.}(2023)\citenamefont {Wen}, \citenamefont {Gao}, \citenamefont {Zhang}, \citenamefont {Tu}, \citenamefont {Li}, \citenamefont {Du}, \citenamefont {Liu},\ and\ \citenamefont {Li}}]{Wen:23}%
  \BibitemOpen
  \bibfield  {author} {\bibinfo {author} {\bibfnamefont {T.}~\bibnamefont {Wen}}, \bibinfo {author} {\bibfnamefont {S.}~\bibnamefont {Gao}}, \bibinfo {author} {\bibfnamefont {J.}~\bibnamefont {Zhang}}, \bibinfo {author} {\bibfnamefont {J.}~\bibnamefont {Tu}}, \bibinfo {author} {\bibfnamefont {W.}~\bibnamefont {Li}}, \bibinfo {author} {\bibfnamefont {C.}~\bibnamefont {Du}}, \bibinfo {author} {\bibfnamefont {W.}~\bibnamefont {Liu}},\ and\ \bibinfo {author} {\bibfnamefont {Z.}~\bibnamefont {Li}},\ }\bibfield  {title} {\bibinfo {title} {112 orbital angular momentum modes amplification based on a 7\uppercase{RC-EDF} with low differential mode gain},\ }\href {https://doi.org/10.1364/OL.477168} {\bibfield  {journal} {\bibinfo  {journal} {Opt. Lett.}\ }\textbf {\bibinfo {volume} {48}},\ \bibinfo {pages} {105} (\bibinfo {year} {2023})}\BibitemShut {NoStop}%
\bibitem [{\citenamefont {Brunet}\ \emph {et~al.}(2014)\citenamefont {Brunet}, \citenamefont {Vaity}, \citenamefont {Messaddeq}, \citenamefont {LaRochelle},\ and\ \citenamefont {Rusch}}]{Brunet:14}%
  \BibitemOpen
  \bibfield  {author} {\bibinfo {author} {\bibfnamefont {C.}~\bibnamefont {Brunet}}, \bibinfo {author} {\bibfnamefont {P.}~\bibnamefont {Vaity}}, \bibinfo {author} {\bibfnamefont {Y.}~\bibnamefont {Messaddeq}}, \bibinfo {author} {\bibfnamefont {S.}~\bibnamefont {LaRochelle}},\ and\ \bibinfo {author} {\bibfnamefont {L.~A.}\ \bibnamefont {Rusch}},\ }\bibfield  {title} {\bibinfo {title} {Design, fabrication and validation of an \uppercase{OAM} fiber supporting 36 states},\ }\href {https://doi.org/10.1364/OE.22.026117} {\bibfield  {journal} {\bibinfo  {journal} {Opt. Express}\ }\textbf {\bibinfo {volume} {22}},\ \bibinfo {pages} {26117} (\bibinfo {year} {2014})}\BibitemShut {NoStop}%
\bibitem [{\citenamefont {Ou}\ \emph {et~al.}(2022)\citenamefont {Ou}, \citenamefont {Tu}, \citenamefont {Wen}, \citenamefont {Li}, \citenamefont {Gao}, \citenamefont {Du}, \citenamefont {Zhou}, \citenamefont {Zhang}, \citenamefont {Sui}, \citenamefont {Liu},\ and\ \citenamefont {Li}}]{Ou:22}%
  \BibitemOpen
  \bibfield  {author} {\bibinfo {author} {\bibfnamefont {N.}~\bibnamefont {Ou}}, \bibinfo {author} {\bibfnamefont {J.}~\bibnamefont {Tu}}, \bibinfo {author} {\bibfnamefont {T.}~\bibnamefont {Wen}}, \bibinfo {author} {\bibfnamefont {W.}~\bibnamefont {Li}}, \bibinfo {author} {\bibfnamefont {S.}~\bibnamefont {Gao}}, \bibinfo {author} {\bibfnamefont {C.}~\bibnamefont {Du}}, \bibinfo {author} {\bibfnamefont {J.}~\bibnamefont {Zhou}}, \bibinfo {author} {\bibfnamefont {B.}~\bibnamefont {Zhang}}, \bibinfo {author} {\bibfnamefont {Q.}~\bibnamefont {Sui}}, \bibinfo {author} {\bibfnamefont {W.}~\bibnamefont {Liu}},\ and\ \bibinfo {author} {\bibfnamefont {Z.}~\bibnamefont {Li}},\ }\bibfield  {title} {\bibinfo {title} {Amplification of 20 orbital angular momentum modes based on a ring-core \uppercase{Y}b-doped fiber},\ }\href {https://doi.org/10.1364/OE.455187} {\bibfield  {journal} {\bibinfo  {journal} {Opt. Express}\ }\textbf {\bibinfo {volume} {30}},\ \bibinfo {pages} {18939} (\bibinfo {year} {2022})}\BibitemShut
  {NoStop}%
\bibitem [{\citenamefont {Zalesskaia}\ \emph {et~al.}(2024)\citenamefont {Zalesskaia}, \citenamefont {Lei}, \citenamefont {Kazansky}, \citenamefont {Wondraczek}, \citenamefont {Gumenyuk},\ and\ \citenamefont {Filippov}}]{Zalesskaia:24}%
  \BibitemOpen
  \bibfield  {author} {\bibinfo {author} {\bibfnamefont {I.}~\bibnamefont {Zalesskaia}}, \bibinfo {author} {\bibfnamefont {Y.}~\bibnamefont {Lei}}, \bibinfo {author} {\bibfnamefont {P.~G.}\ \bibnamefont {Kazansky}}, \bibinfo {author} {\bibfnamefont {K.}~\bibnamefont {Wondraczek}}, \bibinfo {author} {\bibfnamefont {R.}~\bibnamefont {Gumenyuk}},\ and\ \bibinfo {author} {\bibfnamefont {V.}~\bibnamefont {Filippov}},\ }\bibfield  {title} {\bibinfo {title} {Double-clad ytterbium-doped tapered fiber with circular birefringence as a gain medium for structured light},\ }\href {https://doi.org/10.1364/OL.506083} {\bibfield  {journal} {\bibinfo  {journal} {Opt. Lett.}\ }\textbf {\bibinfo {volume} {49}},\ \bibinfo {pages} {270} (\bibinfo {year} {2024})}\BibitemShut {NoStop}%
\bibitem [{\citenamefont {Leonhardt}\ and\ \citenamefont {Philbin}(2010)}]{trafoOptics}%
  \BibitemOpen
  \bibfield  {author} {\bibinfo {author} {\bibfnamefont {U.}~\bibnamefont {Leonhardt}}\ and\ \bibinfo {author} {\bibfnamefont {T.}~\bibnamefont {Philbin}},\ }\href@noop {} {\emph {\bibinfo {title} {Geometry of Light: the Science of Invisibility}}}\ (\bibinfo  {publisher} {Dover},\ \bibinfo {year} {2010})\BibitemShut {NoStop}%
\bibitem [{\citenamefont {Zolla}\ \emph {et~al.}(2005)\citenamefont {Zolla}, \citenamefont {Renversez}, \citenamefont {Nicolet}, \citenamefont {Guenneau},\ and\ \citenamefont {Felbacq}}]{PCFbook}%
  \BibitemOpen
  \bibinfo {editor} {\bibfnamefont {F.}~\bibnamefont {Zolla}}, \bibinfo {editor} {\bibfnamefont {G.}~\bibnamefont {Renversez}}, \bibinfo {editor} {\bibfnamefont {A.}~\bibnamefont {Nicolet}}, \bibinfo {editor} {\bibfnamefont {S.}~\bibnamefont {Guenneau}},\ and\ \bibinfo {editor} {\bibfnamefont {D.}~\bibnamefont {Felbacq}},\ eds.,\ \href@noop {} {\emph {\bibinfo {title} {Foundations of photonic crystal fibers}}}\ (\bibinfo  {publisher} {World Scientific},\ \bibinfo {year} {2005})\BibitemShut {NoStop}%
\bibitem [{\citenamefont {Asgharzadeh~B.}\ \emph {et~al.}(2025)\citenamefont {Asgharzadeh~B.}, \citenamefont {Gumenyuk},\ and\ \citenamefont {Ornigotti}}]{HRMarxiv}%
  \BibitemOpen
  \bibfield  {author} {\bibinfo {author} {\bibfnamefont {H.}~\bibnamefont {Asgharzadeh~B.}}, \bibinfo {author} {\bibfnamefont {R.}~\bibnamefont {Gumenyuk}},\ and\ \bibinfo {author} {\bibfnamefont {M.}~\bibnamefont {Ornigotti}},\ }\bibfield  {title} {\bibinfo {title} {Optical vortex dynamics in non-uniform twisted ring-core fibers},\ }\href@noop {} {\bibfield  {journal} {\bibinfo  {journal} {arXiv:2509.26078}\ } (\bibinfo {year} {2025})}\BibitemShut {NoStop}%
\bibitem [{\citenamefont {Vaity}\ \emph {et~al.}(2013)\citenamefont {Vaity}, \citenamefont {Banerji},\ and\ \citenamefont {Singh}}]{VAITY20131154}%
  \BibitemOpen
  \bibfield  {author} {\bibinfo {author} {\bibfnamefont {P.}~\bibnamefont {Vaity}}, \bibinfo {author} {\bibfnamefont {J.}~\bibnamefont {Banerji}},\ and\ \bibinfo {author} {\bibfnamefont {R.}~\bibnamefont {Singh}},\ }\bibfield  {title} {\bibinfo {title} {Measuring the topological charge of an optical vortex by using a tilted convex lens},\ }\href@noop {} {\bibfield  {journal} {\bibinfo  {journal} {Physics Letters A}\ }\textbf {\bibinfo {volume} {377}},\ \bibinfo {pages} {1154} (\bibinfo {year} {2013})}\BibitemShut {NoStop}%
\bibitem [{\citenamefont {Goodman}\ and\ \citenamefont {Lawrence}(1967)}]{Goodman1967}%
  \BibitemOpen
  \bibfield  {author} {\bibinfo {author} {\bibfnamefont {J.~W.}\ \bibnamefont {Goodman}}\ and\ \bibinfo {author} {\bibfnamefont {R.~W.}\ \bibnamefont {Lawrence}},\ }\bibfield  {title} {\bibinfo {title} {Digital image formation from electronically detected holograms},\ }\href {https://doi.org/10.1063/1.1755025} {\bibfield  {journal} {\bibinfo  {journal} {Applied Physics Letters}\ }\textbf {\bibinfo {volume} {11}},\ \bibinfo {pages} {77} (\bibinfo {year} {1967})}\BibitemShut {NoStop}%
\bibitem [{\citenamefont {Verrier}\ and\ \citenamefont {Atlan}(2011)}]{Off-axis:11}%
  \BibitemOpen
  \bibfield  {author} {\bibinfo {author} {\bibfnamefont {N.}~\bibnamefont {Verrier}}\ and\ \bibinfo {author} {\bibfnamefont {M.}~\bibnamefont {Atlan}},\ }\bibfield  {title} {\bibinfo {title} {Off-axis digital hologram reconstruction: some practical considerations},\ }\href {https://doi.org/10.1364/AO.50.00H136} {\bibfield  {journal} {\bibinfo  {journal} {Appl. Opt.}\ }\textbf {\bibinfo {volume} {50}},\ \bibinfo {pages} {H136} (\bibinfo {year} {2011})}\BibitemShut {NoStop}%
\bibitem [{\citenamefont {M\"{o}ller}\ \emph {et~al.}(2023)\citenamefont {M\"{o}ller}, \citenamefont {Palma-Vega}, \citenamefont {Grimm}, \citenamefont {H\"{a}ssner}, \citenamefont {Kuhn}, \citenamefont {Nold}, \citenamefont {Haarlammert}, \citenamefont {Walbaum},\ and\ \citenamefont {Schreiber}}]{Moller:23}%
  \BibitemOpen
  \bibfield  {author} {\bibinfo {author} {\bibfnamefont {F.}~\bibnamefont {M\"{o}ller}}, \bibinfo {author} {\bibfnamefont {G.}~\bibnamefont {Palma-Vega}}, \bibinfo {author} {\bibfnamefont {F.}~\bibnamefont {Grimm}}, \bibinfo {author} {\bibfnamefont {D.}~\bibnamefont {H\"{a}ssner}}, \bibinfo {author} {\bibfnamefont {S.}~\bibnamefont {Kuhn}}, \bibinfo {author} {\bibfnamefont {J.}~\bibnamefont {Nold}}, \bibinfo {author} {\bibfnamefont {N.}~\bibnamefont {Haarlammert}}, \bibinfo {author} {\bibfnamefont {T.}~\bibnamefont {Walbaum}},\ and\ \bibinfo {author} {\bibfnamefont {T.}~\bibnamefont {Schreiber}},\ }\bibfield  {title} {\bibinfo {title} {Polarization-resolved mode evolution in \uppercase{tmi}-limited \uppercase{Y}b-doped fiber amplifiers using a novel high-speed stokes polarimeter},\ }\href {https://doi.org/10.1364/OE.505716} {\bibfield  {journal} {\bibinfo  {journal} {Opt. Express}\ }\textbf {\bibinfo {volume} {31}},\ \bibinfo {pages} {44486} (\bibinfo {year} {2023})}\BibitemShut {NoStop}%
\bibitem [{\citenamefont {Ulrich}\ \emph {et~al.}(1980)\citenamefont {Ulrich}, \citenamefont {Rashleigh},\ and\ \citenamefont {Eickhoff}}]{Ulrich:80}%
  \BibitemOpen
  \bibfield  {author} {\bibinfo {author} {\bibfnamefont {R.}~\bibnamefont {Ulrich}}, \bibinfo {author} {\bibfnamefont {S.~C.}\ \bibnamefont {Rashleigh}},\ and\ \bibinfo {author} {\bibfnamefont {W.}~\bibnamefont {Eickhoff}},\ }\bibfield  {title} {\bibinfo {title} {Bending-induced birefringence in single-mode fibers},\ }\href {https://doi.org/10.1364/OL.5.000273} {\bibfield  {journal} {\bibinfo  {journal} {Opt. Lett.}\ }\textbf {\bibinfo {volume} {5}},\ \bibinfo {pages} {273} (\bibinfo {year} {1980})}\BibitemShut {NoStop}%
\end{thebibliography}

% Full bibliography added automatically for Optics Letters submissions; the following line will simply be ignored if submitting to other journals.
% Note that this extra page will not count against page length

\end{document}